\documentclass[11pt,a4paper]{article}
\pdfoutput=1

\usepackage{jheppub}
\usepackage{amsfonts}
\usepackage{amsmath,amssymb}
\usepackage{amsthm}
\usepackage{graphicx}
\usepackage{wrapfig}

\newcommand{\n}{\noindent}

\newcommand{\ii}{\textrm{i}}
\newcommand{\dd}{\textrm{d}}
\newcommand{\e}{\textrm{e}}

\newcommand{\ga}{\alpha}
\newcommand{\gb}{\beta}
\renewcommand{\gg}{\gamma}
\newcommand{\gd}{\delta}
\newcommand{\gep}{\epsilon}

\newcommand{\gth}{\theta}

\newcommand{\gl}{\lambda}

\newcommand{\gs}{\sigma}

\newcommand{\Gg}{\Gamma}
\newcommand{\Gd}{\Delta}

\newcommand{\Wbar}{\overline{W}}

\newcommand{\cC}{\mathcal C}
\newcommand{\cD}{\mathcal D}

\newcommand{\cL}{\mathcal L}

\newcommand{\cN}{\mathcal N}
\newcommand{\cO}{\mathcal O}
\newcommand{\cP}{\mathcal P}

\newcommand{\cR}{\mathcal R}

\newcommand{\Tr}{\mbox{Tr}}

\newcommand{\be}{\begin{equation}}
\newcommand{\bea}{\begin{eqnarray}}
\newcommand{\ee}{\end{equation}}
\newcommand{\eea}{\end{eqnarray}}
\renewcommand{\d}{\partial}

\newcommand{\half}{\frac{1}{2}}

\newcommand{\ov}[1]{\frac{1}{#1}}

\theoremstyle{definition}

\title{Resumming planar diagrams for the $\cN=6$ ABJM cusped Wilson loop in light-cone gauge}

\author{Daniele Marmiroli}
\affiliation{Dipartimento di Fisica, Universit\`a di Parma and INFN Gruppo Collegato di Parma, \\ Viale G.P. Usberti 7/A, 43100 Parma, Italy}
\emailAdd{daniele.marmiroli@fis.unipr.it}

\abstract{We analyse a light-like cusped Wilson loop in $\cN=6$ superconformal Chern-Simons theory at both weak and strong coupling in light-cone gauge. At the second order in the 't~Hooft coupling $\gl$ the correct cusp anomalous dimension $\Gg_{\rm cusp}=-\frac{\phi}{2}\gl^2$ is recovered through a deformation of the contour that takes both rays of the cusp slightly off of the light-cone. The strong coupling behaviour is addressed by means of the Bethe-Salpeter equation for ladders of tree-level gauge propagators and ladders of one-loop corrected gauge propagators. It turns out that, as might be expected, the contribution of Chern-Simons tree-level propagators is insensitive of the cusp angle $\phi$. On the other hand, corrected propagators lead to an exponential large $\gl$ behaviour $\Gg_{\rm cusp} \sim \exp\sqrt{\gl\phi}$ which, though, disagrees with the AdS/CFT predictions in the power of $\phi$.}


\begin{document}
\maketitle

\newpage
\section{Introduction}
\label{sec:intro}

A lot of interest has recently arouse about cusped Wilson loops and their supersymmetric deformations at both weak coupling, from gauge theories computations, and at strong coupling, by means of AdS/CFT. Cusped loops are interesting objects for various reasons. In supersymmetric theories, they describe the trajectory of W-bosons moving on a straight line, suddenly changing direction by an angle $\phi$ and then moving on a straight line again. Such loops are multiplicatively renormalisable in Euclidean spacetime (or in Minkowsky spacetime, provided they lie away from the lightcone) \cite{Korchemsky:1987wg}. The logarithmic divergence due to the emission of soft fields defines the cusp anomalous dimension through the relation

\be
\log\left< W \right> \sim -\Gg_{\rm cusp}(\gl,\phi)\log\frac{L}{\gep} 
\ee

being $L,\gep$ respectively IR and UV cutoffs on the contour and $\gl$ the 't~Hooft coupling constant. In the planar limit, $\Gg_{\rm cusp}$ essentially dominates the IR divergence of scattering amplitudes of massive particles providing a factor of $\Gg_{\rm cusp}(\gl,\phi_i)$ for each couple of particles at boost angles $\phi_i$. Analytically continuing $\phi\to\ii\varphi$, the $\varphi\to\infty$ limit of the cusp anomalous dimension governs the IR divergence of scattering amplitudes of massless particles \cite{Korchemsky:1987wg,Korchemsky:1991zp,Korchemsky:1988si}, in such limit 

\be
\label{eq:gamma-definition}
\Gg_{\rm cusp}(\gl,\ii\varphi)=\frac{\varphi}{2}\Gg_{\rm cusp}(\gl)
\ee

In $\cN=4$ SYM $\Gg_{\rm cusp}(\gl)$, which is also known as cusp anomalous dimension by an abuse of notation, is related to the anomalous dimension of twist-two operators with large spin J \cite{Balitsky:1987bk,Korchemsky:1992xv,Craigie:1980qs}, and in turn to the leading Regge trajectory of a closed string propagating in $AdS_5\times S^5$ with large angular momentum in $AdS$ \cite{Gubser:2002tv,Frolov:2002av}, which predicts at strong coupling 

\be
\Gg_{\rm cusp}(\gl)=\frac{\sqrt{\gl}}{\pi} -\frac{3\log 2}{\pi} + \cO(1/\sqrt{\gl})
\ee

The $\sqrt{\gl}$ behaviour was then confirmed from the gauge theory side using integrability in \cite{Beisert:2006ez}.
Making direct avail of the AdS/CFT correspondence, the cusped loop can be also calculated at strong coupling  \cite{Rey:1998ik,Maldacena:1998im,Berenstein:1998ij,Drukker:1999zq}
 through the duality between the supersymmetric Maldacena-Wilson loop operator and an open string whose endpoints are bounded to move along the loop contour. The properly regularised area of the string solution minimising the Nambu-Goto action, and with the boundary conditions above, determines the large $\gl$ behaviour of the Wilson loop. Coupling to both gauge fields and  scalars, the supersymmetric loop operator of $\cN=4$ SYM 

\be
\label{eq:gencuspN=4}
W\sim \Tr\cP \e^{\ii\int A_\mu dx^\mu +\int \Phi_m n^m |dx|}
\ee

allows the introduction of a second parameter, the angle $\gth$ between the two directions $\vec n$ and $\vec n'$  in $S^5$ that define different couplings on different sides of the cusped contour. In particular, such operator coupled to a cusp in Euclidean spacetime has provided a locally supersymmetric observable that, through a conformal mapping, interpolates smoothly between the cusp anomalous dimension in $\mathbb{R}^{1,3}$ and the static $W\Wbar$-pair potential in $S^3\times\mathbb{R}$. Such generalised cusp anomalous dimension $\Gg_{\rm cusp}(\gl,\phi,\gth)$ has been extensively studied by direct perturbative computations \cite{Drukker:2011za,Correa:2012nk}, localization and integrability \cite{Correa:2012at,Drukker:2012de,Correa:2012hh}, and through AdS/CFT \cite{Drukker:2011za}, disclosing a variety of new features and an infinite family of BPS configurations of the Zarembo kind \cite{Zarembo:2002an} whenever $|\phi|=|\gth|$.\\

The existence and properties of the generalised cusp anomalous dimension seem rather general aspects of supersymmetric gauge theories, not confined to the realm of $\cN=4$ SYM. The analogous of (\ref{eq:gencuspN=4}) for $\cN=6$ superconformal Chern-Simons theories of the ABJM kind \cite{Aharony:2008ug,Aharony:2008gk} has been constructed and studied in \cite{Griguolo:2012iq} on the gauge theory side, and in \cite{Forini:2012bb} on the dual $AdS_4\times \mathbb{C}P^3$ superstring side. The construction follows naturally from a deformation of the $\half$ BPS operator of \cite{Drukker:2009hy}, that in turn includes, in a superconnection structure, the coupling of gauge, scalars and fermion fields to the contour. Besides sharing most of its general properties with its four-dimensional counterpart, this operator displays distinctive features. \\

But still, the subject of Wilson loops in SCS theories is a rather unexplored one. Polygonal light-like loops have been considered within the framework of the scattering amplitudes/Wilson loops duality \cite{Henn:2010ps} and have been shown to validate the equivalence at two-loops \cite{Chen:2011vv,Bianchi:2011dg,Bianchi:2011fc}. Quite interestingly, supersymmetric localisation techniques have been successful in reducing the circular loop path-integral to a matrix integral \cite{Kapustin:2009kz}, providing a first example of weak-strong coupling interpolating function for this class of theories. Moreover, several aspects of the partition function of the ABJM theory on the three-sphere have been analysed through matrix models and topological strings techniques \cite{Marino:2009jd,Drukker:2010nc,Drukker:2011zy,Marino:2011eh}, shedding some light on the strong coupling region of SCS theories. More recently, two new and more general classes of supersymmetric loops have been discovered \cite{Cardinali:2012ru}, they represent natural candidates for localisation techniques and may provide suitable observables for a thorough investigation of the AdS$_4$/CFT$_3$ correspondence.\\

An alternative strategy to obtain some information about the large 't~Hooft coupling limit of gauge theories is through resummation of perturbative series. The use of Bethe-Salpeter equations to re-sum  certain classes of Feynman diagrams dates back to early days of studies of hadronic bound states \cite{Salpeter:1951sz}. In the context of supersymmetric gauge theories, the Bethe-Salpeter equation for ladder diagrams contributing to the $\cN=4$ rectangular Wilson loop was derived and solved in \cite{Erickson:1999qv}, and further exploited in \cite{Erickson:2000af}. It predicts the $\exp\sqrt{\gl}$ behaviour of a circular loop and of a couple of anti-parallel lines as it is obtained from semiclassical string computations. On the other hand, it has also been shown that the sum of ladder diagrams does not reproduce the correct behaviour of the cusp anomalous dimension as expected from AdS/CFT, and conjectured that the contribution of gauge field vertices might be crucial in reconstructing the expected strong coupling asymptotics \cite{Makeenko:2006ds}. More recently, a new limit has been identified \cite{Correa:2012nk} in which resummation techniques prove to be particularly fruitful. It is based on the observation that, in the weak coupling expansion of (\ref{eq:gencuspN=4}) diagrams where scalar fields are involved are proportional to $\cos\gth$ and become dominant if the $\gth\to\ii\infty$ limit is properly taken. The full perturbative expansion of the generalised cusp anomalous dimension $\Gg_{\rm cusp}(\gl,\phi,\gth)$ can hence be reorganised in terms of the rescaled coupling $\hat\gl=\e^{\gth}\gl$, then the resummation of ladders of scalar propagators  reproduces the AdS result, including next-to-leading-order, $\cO(\gl)$ corrections \cite{Bykov:2012sc}, once the appropriate limit is take in the string solution.\\

Though, the case of $\cN=6$ SCS theory is somewhat different. The subtle interplay of the fermionic and bosonic sectors suggests that even in this limit, one would have to face the issue of handling fermion-gauge field three-vertices. In this short letter we propose a different strategy. We consider a cusped loop in Minkowski spacetime slightly taken out of the light cone and solve the Bethe-Salpeter equation for ladders of gauge field propagators. In the light-like limit the supersymmetric loop operator reduces to the ordinary one; scalars and fermions contributions, proportional to the moduli of the vectors describing the contour, are suppressed and gauge fields become dominant. Also, we choose an axial gauge fixing for the Chern-Simons fields, which in our case appears as a natural choice from the perturbative point of view. Indeed, Chern-Simons theory in light-like gauge has been extensively studied in relation to knot theory
 \cite{Frohlich:1989gr,AlvarezGaume:1989wk,Alvarez:1991sx,Labastida:1997uw} (and many others) and is known to deeply simplify the perturbative series expansion. Furthermore, light-like axial quantisation has been successfully employed in other contexts, among others the investigation of the cusp anomalous dimension in SYM theory \cite{Bassetto:1993xd} and as an effective computational tool for hadronic bound-state energies in QCD \cite{Brodsky:1997de}. By this choice we are able to reproduce known results for the cusp anomalous dimension at order $\gl^2$. We then formulate the Bethe-Salpeter equation for ladders of tree-level CS propagators and ladders of one-loop corrected propagator. Solving these equations and extrapolating to strong coupling we find a $\exp\sqrt{\gl}$ behaviour for our deformed loop, but with the wrong coefficient according to the AdS/CFT predictions. As a side result we present in Appendix the computation of the one-loop corrected CS propagator in ABJ theories in light-cone gauge.


\section{Cusped Wilson loop in light-cone gauge}
\label{sec:cuspinlightcone}

The $\half$ BPS Wilson loop operator for $\cN=6$  $U(N)\times U(M)$ superconformal Chern-Simons theory proposed in \cite{Drukker:2006zk} is relies on the lift of the gauge group to the supergroup $U(N|M)$, for which a superconnection $\cL$ is constructed based on a suitable extension of the super Wilson loop of $\cN=4$ SYM

\be
\label{superconnection}
 \mathcal{L}(\tau) \equiv \begin{pmatrix}
A_{\mu} \dot x^{\mu}+\frac{2 \pi}{k} |\dot x| M^I_J C_{I}\bar C^J
&\sqrt{\frac{2\pi}{k}}  |\dot x | \eta_{I}\bar\psi^{I}\\
\sqrt{\frac{2\pi}{k}}   |\dot x | \psi_{I}\bar{\eta}^{I} &
\widehat{A}_{\mu} \dot x^{\mu}+\frac{2 \pi}{k} |\dot x| \widehat M^I_J\bar C^J C_{I}
\end{pmatrix}
\ee

where the requirement of supersymmetry invariance along a straight line restricts the fermionic and bosonic couplings $\eta_I^\ga,\, \bar \eta_\ga^I$ and $M^I_J,\, \widehat M^I_J$ to be of the form\footnote{The requirement that $\gd_{\rm susy}\cL(\tau)=0$ for a straight line, actually fixes $\eta=\bar\eta=0$. This means that only purely bosonic operators are allowed, which are at most $\frac{1}{6}$ BPS \cite{Drukker:2008zx,Rey:2008bh}. In order to obtain more supersymmetric operators one has to admit the looser condition $\delta_{\rm susy}\mathcal{L}(\tau)=\mathfrak{D}_{\tau} G\equiv\partial_{\tau} G+ i\{ \mathcal{L},G]$.}

\be
\label{cc}
\eta_{I}^{\alpha}=n_{I} \eta^{\alpha},\ \ \ \   \bar\eta^{I}_{\alpha}=\bar n^{I} \bar\eta_{\alpha},\ \ \ \ 
M_{J}^{\ \ I}=\delta^{I}_{ J}-2 n_{J}  \bar n^{I},\ \ \ \ 
\widehat M_{J}^{\ \ I}=\delta^{I}_{J}-2 n_{J} \bar n^{I}
\ee

Here Greek indices $\ga=1,2$ are spinor indices, $I,J=1,2,3,4$ are $R-$symmetry indices. These conditions brake the $SU(4)$ $R-$symmetry to $SU(3)\times U(1)$, the two complex conjugated vectors $n,\bar n$ then identify the direction in the internal space that is preserved by the action of the $SU(3)$ subgroup. It was conveniently chose in \cite{Griguolo:2012iq} to couple the left and right hand sides of the cusp to fermion and scalars such that 

\be
n_{1I}\bar n_2^I = \cos\frac{\vartheta}{2} \qquad \eta_1^\ga \bar\eta_{2\ga}= \cos\frac{\phi}{2} 
\ee

where $\phi$ is the cusp angle and $\vartheta$ is some angle in the internal space. Hence, diagrams involving the exchange of either scalars or fermions gain a factor of $\cos\frac{\vartheta}{2}$ for each line going from one side to the other side. Following \cite{Correa:2012nk}, one could in principle introduce a new scaling limit in which $\vartheta\to\ii\infty$ and reorganise the perturbative expansion in terms of the rescaled coupling constant $\hat\gl=\e^\vartheta \gl$. Clearly, graphs involving only fermion and scalars exchanges become dominant in this limit, but as was shown in \cite{Griguolo:2012iq}, the cancellation between scalars and fermions would force to consider the fermion-gauge field vertex already at two-loops, and the scalar-gauge vertex and Yukawa couplings at higher loops. We would like to stress that, at strong coupling, the contribution of the fermion-gauge vertex is essential already at leading order in $\hat\gl$ because, as mentioned in the following (\ref{eq:one-fermion}), the contribution of the only single exchange diagram is insensitive of the cusp angle. \\

One could ask what would happen in the opposite case, meaning the one in which gauge fields are dominant. This is achieved by either sending $\vartheta\to\pi$ or considering a strictly light-like contour. In the latter case fermions and scalars decouple from the contour and one is left with the pure gauge Wilson loop  in Lorentz signature

\be
\label{eq:loopoperator}
\left< W(\cC) \right> = \frac{1}{{\rm dim}\cR + {\rm dim \hat\cR}} \int \cD[{\rm fields}]\e^{\ii S_{ABJ}} \left[ \Tr_\cR \cP \, \e^{\ii \int_\cC A_m \dd x^m} +\Tr_{\hat\cR} \cP \, \e^{\ii \int_\cC \hat A_n \dd x^n} \right]
\ee  

\begin{wrapfigure}[18]{l}{60mm}
\includegraphics[width=.30\textwidth]{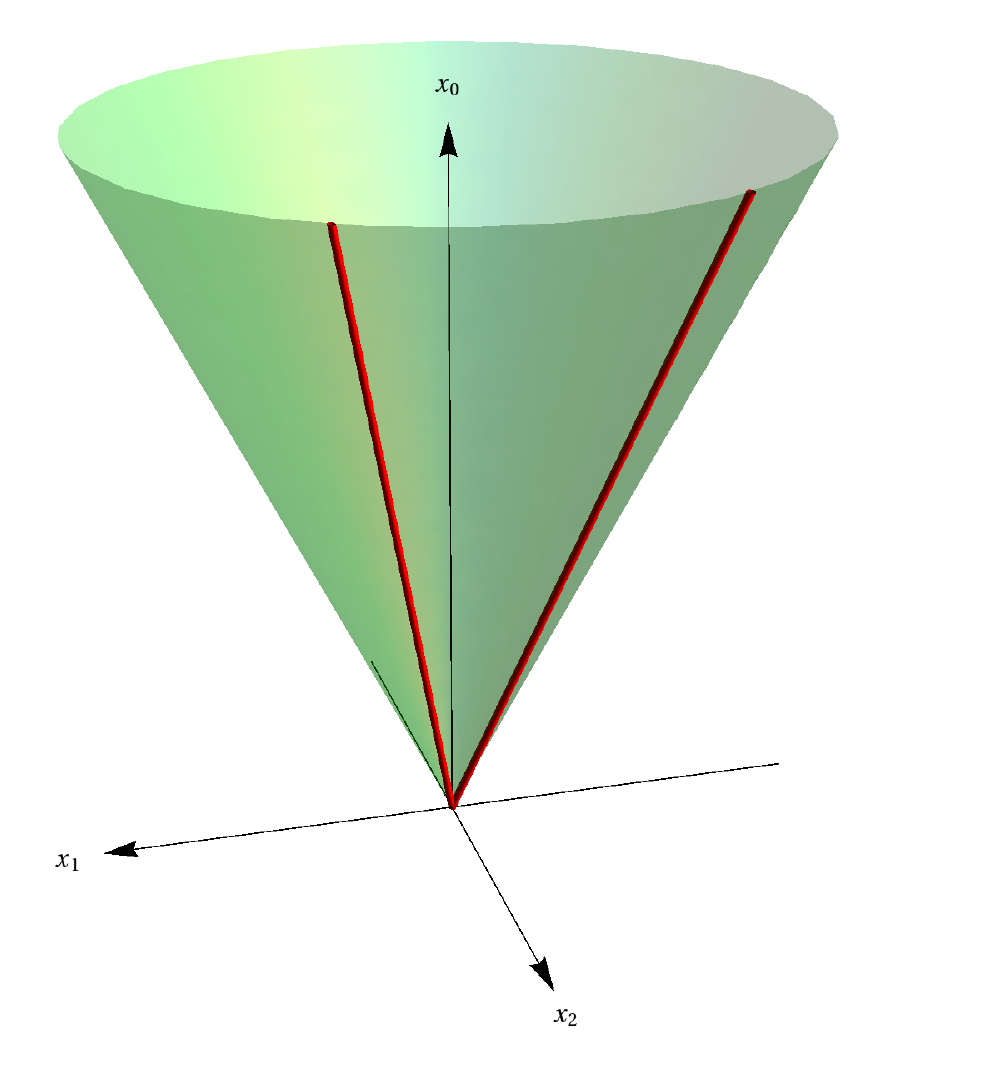}
\caption{\label{cusptowedge}The two semi-infinite lines forming the light-like cusp lie on the same light-cone at some geometric angle $0<\gth<\pi$, not to be confused with the cusp angle $\phi$.}
\end{wrapfigure}

where the trace is in some representation $\cR$ of $SU(N)$ and its conjugate $\hat\cR$, in the following the fundamental and anti-fundamental.
Note that this is not the only gauge invariant loop operator one can choose, indeed we could compute the expectation value of any combination of the traces of two holonomies in any two representations, see \cite{Drukker:2008zx} for a discussion about this point. In particular the choice of the relative sign is not in principle determined by the requirement of gauge invariance. On the other hand we want to consider the legitimate light-like limit of (\ref{superconnection}), and this fact uniquely fixes the form of (\ref{eq:loopoperator}) we have considered. This said, let us consider a cusped Wilson loop in Minkowski $\mathbb{R}^{2,1}$ spacetime, {\it i.e.} a couple of semi-infinite Wilson lines
lying on the same light-cone and joined together at the vertex
of the cone. Such a contour can be parametrised by a couple of vectors $\dot x_1^\mu, \dot x_2^\mu$

\be
\label{eq:cuspparametrisation}
\cC(s) = \left\{
\begin{array}{cc}
 \dot x_1 s=\{-s, s \cos\gth, -s \sin\gth\} & \quad s<0\\
 \dot x_2 s=\{s, s \cos\gth, s \sin\gth\} & \quad s>0
\end{array} \right.
\ee

where we still used Lorentz indices $\mu=0,1,2$. The angle $\gth\in [0,\pi]$ parametrises the geometrical separation between the two legs of the cusp, and is not to be confused with the velocity $\phi$, also known as cusp angle, nor with the internal angle $\vartheta$ introduced before and that will be irrelevant in the following. Indeed, for any two vectors $p^\mu$ and $q^\mu$, in Euclidean spacetime one has

\be
\frac{p\cdot q}{|p||q|}= \cos\gth
\ee

for some angle $\theta$ which is finite and describes an Euclidean angle. On the other hand, on the Minkowski side, the proper light-like cusp is described by two null vectors $|p|=|q|=0$ and the relation above is analytically continued to

\be
\frac{p\cdot q}{|p||q|}= \cosh \phi
\ee

for some infinite angle $\phi$. To make the correspondence clearer, let us slightly take the contour $\cC$ out of the light cone through the deformation parameter $a\sim 0$

\be
\label{eq:cuspdeformed}
\cC_{\rm def}(s) = \left\{
\begin{array}{cc}
 \{-s, s \cos\gth, -s \sin\gth+a\} & \quad s<0\\
 \{s, s \cos\gth, s \sin\gth+a\} & \quad s>0
\end{array} \right.
\ee

Thus, being $\dot x_1 \cdot \dot x_2= -2\cos^2\gth -a^2$ and $|\dot x_1|^2=|\dot x_2|^2=2a\sin\gth$, as we approach the light-cone sending $a\to 0$ we have the relation

\be
\label{eq:aphimap}
\phi = -\log\,a +\log \frac{\sin^2\gth -1}{\sin\gth}
\ee

which makes clear that the large $\phi$ expansion singles out $\log\,a$ contributions whereas $\gth$ dependent ones are subleading. Otherwise stated, the leading contribution to $W(\cC_{\rm def})$ does not know about the geometry of the cusp (away from $\gth=0$ where the cusp degenerates), it  does only know that it lies on a lightcone. \\

The path-ordered exponential (\ref{eq:loopoperator}) admits the weak coupling expansion (we will use light-cone coordinates $x^m$ and light-cone indices $m=+,-,T$ from now on, useful definitions and conventions are in Appendix \ref{sec:csaction})

\be
\label{eq:loopexpansion}
\begin{split}
\cP \, \e^{\ii \int_\cC \dd s_i\, A_m \dot x_i^m } =& \mathbb{I} + \ii \int_\cC \dd s_i \,\dot x_i^m A_m - \int_\cC \dd s_i \,\int_{s_j<s_i} \dd s_j \, \dot x_i^m \dot x_j^n A^m A^n\\
& -\ii \int_\cC \dd s_i \,\int_{s_j<s_i} \dd s_j\, \int_{s_k<s_j} \dd s_k\, \dot x_i^m \dot x_j^n \dot x_k^r A_m A_n A_r\\
& + \int_\cC \dd s_i \,\int_{s_j<s_i} \dd s_j\, \int_{s_k<s_j} \dd s_k\,  \int_{s_l<s_k} \dd s_l\, \dot x_i^m \dot x_j^n \dot x_k^r \dot x_l^s A_m A_n A_r A_s\\
&+ \dots
\end{split} 
\ee

and analogously for the $\hat A$ gauge field. The first few orders of the perturbative expansion get substantially simplified by chosing a light-cone gauge fixing for both the gauge fields \cite{Frohlich:1989gr,AlvarezGaume:1989wk}

\be
A_-=A_0 - A_1=0 \qquad \hat A_-=\hat A_0 - \hat A_1=0
\ee

This way the three-vertex in the Chern-Simons action becomes identically zero and ghosts decouple from the action, leaving a free-field Lagrangian

\be
\begin{split}
& S_{CS} = \frac{k}{4\pi}\int \dd^3x\,\gep^{mn}A_m^a \d_{-} A_n^a\\
& S_{gh} = \int \dd^3x\,\bar c^a \d_{-} c^a 
\end{split}
\ee

The matter part of the ABJ Lagrangean is left basically untouched, besides the fact that the covariant derivative $D_-$ becomes the standard derivative $\d_-$.
The perturbative expansion of pure Chern-Simons theory in this gauge is only populated by ladder graphs, and interactions in ABJ theory arise only at higher orders through loops of scalar and fermions as effective vertices coupled to gauge fields.  More details can be found in Appendix \ref{sec:csaction}.


\section{Light-like versus deformed cusp at two-loops}
\label{sec:loops}

\subsection{The strictly light-like case}
\label{sec:strictlight}

The tree-level propagator of Chern-Simons fields in light-cone gauge was computed in \cite{Frohlich:1989gr} long ago (see Appendix \ref{sec:csaction} also)

\be
\left< (A_{ij})_m(x) (A_{kl})_n^b(y) \right> = -\frac{1}{k}\gep_{mn}\gd_{il}\gd_{jk} \frac{\gd(x_T-y_T)}{[x_- -y_-]}
\ee

and similarly for the second gauge group

\be
\left< (\hat A_{\bar i \bar j})_m(x) (\hat A_{\bar k \bar l})_n(y) \right> = \frac{1}{k}\gep_{mn}\gd_{\bar i \bar l} \gd_{\bar j \bar k} \frac{\gd(x_T-y_T)}{[x_- -y_-]}
\ee

where we have explicitly written group indices $i,j,..=1,..N;\,\bar i,\bar j,.. =1,..M $. Note that the dynamics is restricted to a two dimensional subspace $\mathbb{R}^{1,1}$ of the three-dimensional Minkowski space, hence the propagators above can also be written as the light-cone derivative of the two-dimensional Feynman propagator

\be
G_{mn}^{ab}(x)=-\ii\gd^{ab} \gep_{mn}\d_+ \Gd_{F}(x_+,x_-)\gd(x_T)
\ee

There is just one contribution to compute at one-loop order, the single gluon exchange, which reads

\be
\begin{split}
\left<W^{(1)}_{A}\right> =& -\frac{1}{N+M} \int_\cC \dd s_1 \,\int_{s_2<s_1} \dd s_2 \, \dot x_1^m \dot x_2^n \left< A_m(s_1) A_n(s_2) \right>\\
=& \frac{1}{N+M}\frac{N^2}{k} \int_\cC \dd s_1 \,\int_{s_2<s_1} \dd s_2 \, \frac{\dot x_1^+ \dot x_2^T - \dot x_1^T \dot x_2^+}{\dot x_1^+ s_1-\dot x_2^+ s_2}\gd(\dot x_1^T s_1 -\dot x_2^T s_2)\\
\end{split}
\ee

and evidently vanishes if both ends of the propagator are stuck to the same leg of the cusp. The only region onto which it has to be integrated is

\be
\begin{split}
\label{eq:onegluon}
& \frac{1}{N+M}\frac{N^2}{k} \int_{-L}^{-\gep} \dd s_1 \,\int_\gep^L \dd s_2 \, \frac{\dot x_1^+ \dot x_2^T - \dot x_1^T \dot x_2^+}{\dot x_1^+ s_1-\dot x_2^+ s_2}\gd(\dot x_1^T s_1 -\dot x_2^T s_2)\\
=& \frac{1}{N+M}\frac{N^2}{k} \int_\gep^L \dd s_2 \, \frac{ {\rm sign}\,\dot x_1^T }{s_2}\Theta(\tilde s_1 + L) \Theta(-\gep-\tilde s_1)\\
=&  \frac{1}{N+M}\frac{N^2}{k} \log\frac{L}{\gep}
\end{split}
\ee

where $\tilde s_1= \frac{\dot x_2^T}{\dot x_1^T} s_2$ is the solution of the delta function constraint and thetas on the second line force the integration solely over the region where such a solution exists. In our case such constraints are trivial as $\dot x_1^T = \dot x_2^T = \sin\phi >0$. There is a totally identical contribution arising from the $\hat A$ field, the only difference being a minus sign in the propagator and a different group factor, namely

\be
\left<W^{(1)}_{\hat A}\right>=-\frac{1}{N+M}\frac{M^2}{k} \log\frac{L}{\gep}
\ee

From this and from (\ref{eq:loopoperator}) we read the one-loop expectation value of the cusped Wilson loop

\be
\label{eq:Woneloop}
\left< W^{(1)}(\cC) \right> = \frac{N-M}{k} \log\frac{L}{\gep}
\ee

As expected it does not depend on any kinematic detail, for the theory is topological at this order, in particular it does not carry any information about  the cusp angle $\phi$. For this reason it is likely that ladders of CS propagators will not contribute to the cusp anomalous dimension, nor at weak neither at strong coupling, but they will only contribute, eventually, to the renormalisation factor of the infinite contour $\cC$. \\

\begin{figure}[htb]
\begin{center}
\fbox{\includegraphics[width=.15\textwidth]{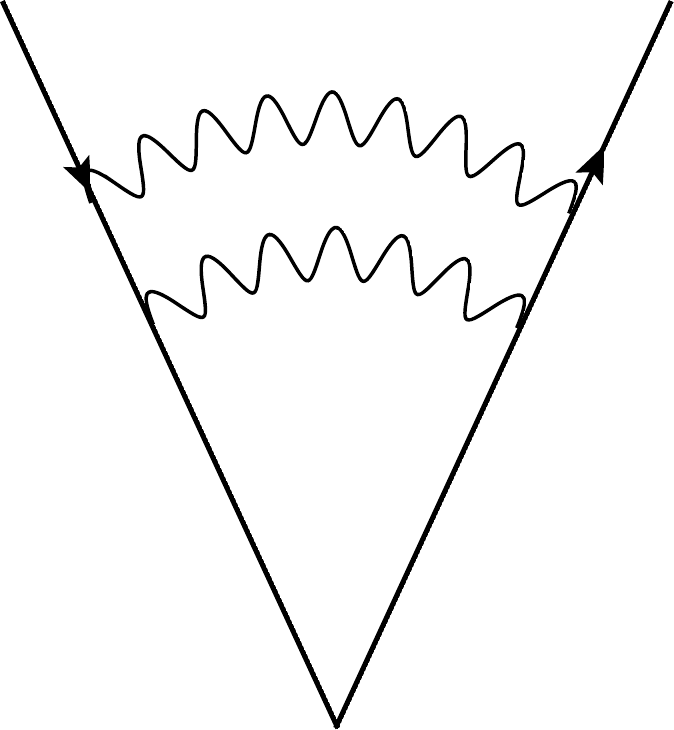}\mbox{(a)}}
\fbox{\includegraphics[width=.15\textwidth]{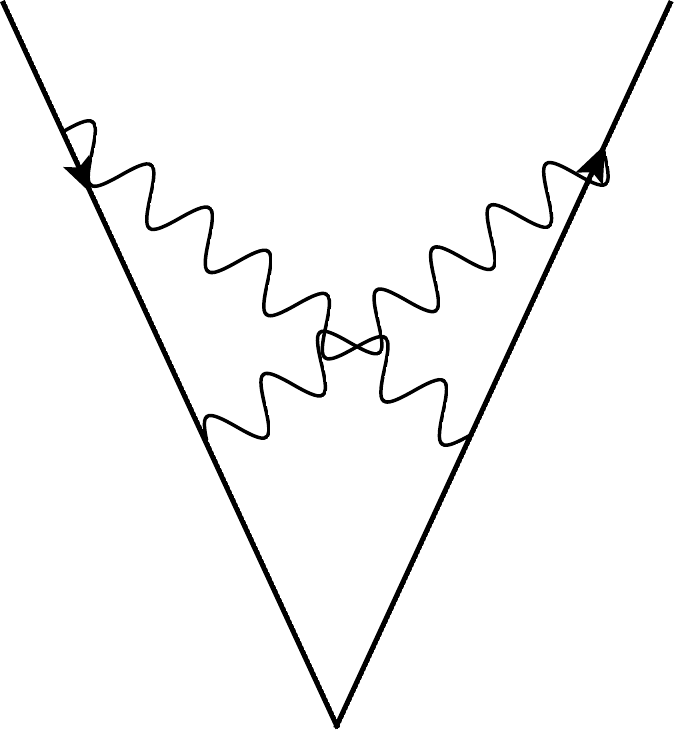}\mbox{(b)}}
\fbox{\includegraphics[width=.15\textwidth]{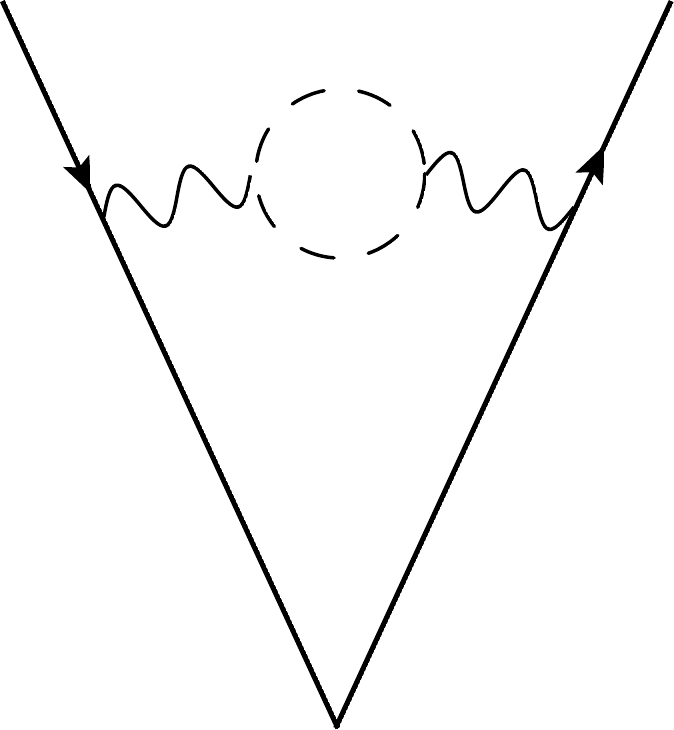}\mbox{(c)}}
\fbox{\includegraphics[width=.15\textwidth]{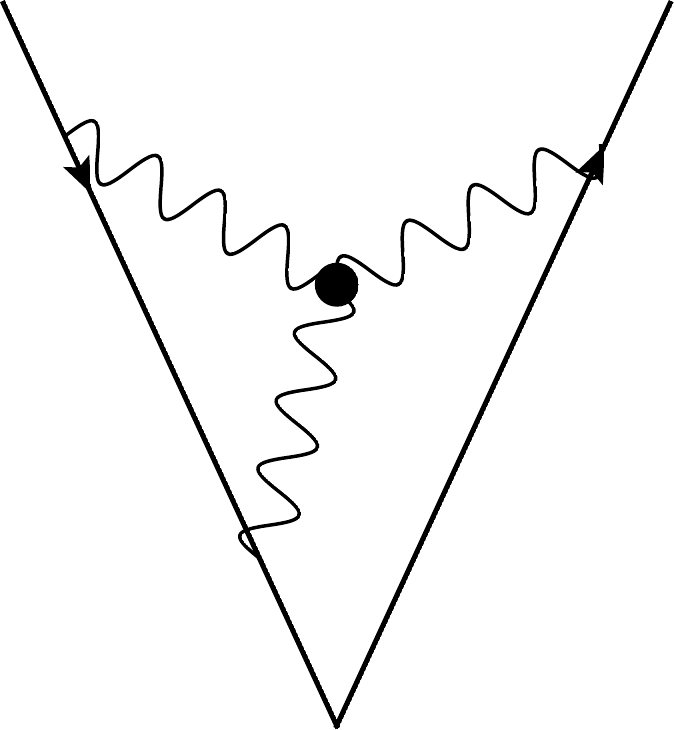}\mbox{(d)}}
\end{center}
\caption[Graphs contributing at two-loops.]{\label{fig:twoloopsgraphs}  Only two kinds of graphs contribute to the two-loops expansion in light-cone gauge: double gluon exchanges (a),(b) and the one-loop corrected gluon propagator (c). The three-gluon vertex totally decouples from the action.}
\end{figure}

Going forward to the two-loops order, the perturbative expansion of pure Chern-Simons theory in light-cone gauge is in principle straightforward, since the gauge field vertex disappears and ghosts decouple from the action, leaving a theory where only propagators are allowed.
Coupling the gauge field to fermionic and scalar matter introduces, at this loop order, only the self energy correction to the CS field, computed in Appendix \ref{sec:corrprop}. Graphs where at least one tree-level propagator is attached with both ends on the same leg of the contour vanish, as in the one-loop case, hence  at two-loops there are only three graphs in light-cone gauge; they are depicted in Figure \ref{fig:twoloopsgraphs}. Opposed to the four-dimensional case, the contribution of (c) on a single leg of the contour differs from zero. The fact that single-leg diagrams actively take part in the perturbative expansion of the Wilson loop is a distinguishing feature of superconformal Chern-Simons theories which was already observed in \cite{Griguolo:2012iq}.

\vspace{10mm}
{\bf Double exchange diagrams}\\

Diagrams with two tree-level propagators, Figure \ref{fig:twoloopsgraphs} (a) and (b), arise from the fourth-order expansion of the loop operator

\be
\label{eq:doubleexchangegraphs}
\int \dd s_i \,\dd s_j\, \dd s_k\, \dd s_l\, \dot x_i^m \dot x_j^n \dot x_k^r \dot x_l^s \bigg(\left< A_m A_n \right> \left< A_r A_s \right> + \left< A_m A_s \right> \left< A_n A_s\right> + \left< A_m A_r \right> \left< A_n A_s\right> \bigg)
\ee

Of the three summands above, the first always contains a single-leg tree-level propagator, and the third produces non-planar diagrams, so only the second summand is to take into account, strictly speaking.
These graphs are analysed in full details in Appendix \ref{sec:on-lightlike} (including non-planar contributions \ref{fig:twoloopsgraphs}(b) ). The contribution of (a) is easily seen to exponentiate the single gauge field exchange

\be
\label{eq:twogluons}
I_{(a)} = \frac{N^3}{2k^2}\left( \log\frac{L}{\gep} \right)^2 + \cO\left(\frac{\gep}{L}\right)
\ee

\vspace{10mm}
{\bf Corrected propagator}
~\\

The second order of  (\ref{eq:loopexpansion}) contracted with two powers of the interaction Lagrangian produces the CS self-energy graph of Figure \ref{fig:twoloopsgraphs} (c)

\be
I_{(c)}=- \int \dd s_i\,\dd s_j \, \dot x_i^m \dot x_j^n \, A_m A_n\,\half \left( \ii\int \dd^3 y \cL_{\rm int}(y) \right)^2 = \half \int \dd s_i\,\dd s_j \, \dot x_i^m \dot x_j^n G_{mn}^{(2)}(x_i,x_j)
\ee 

which  also contributes to the $\gl^2$ order of the Wilson loop.
The one-loop corrected CS propagator in strictly three dimensions in light-cone gauge is computed in  Appendix \ref{sec:corrprop} and reads 

\be
\label{eq:G2}
G^{(2)}_{mn} =  \frac{M}{k^2}D_{mn}\left[ -\frac{x^-}{[x^+]} + \half \frac{x^2}{[x^+]^2} \log\left(-\frac{x^2}{(x^T)^2}\right) \right]
\ee

where the derivatives are w.r.t $x^m = (x_1-x_2)^m$ and act as

\be
D_{mn}= \gep_{mr}\gep_{ln} (\eta^{rl} \d^2 - \d^r \d^l ) =\d^2 \eta_{mn} - \d_m \d_n +\gd_{mn}(2\d^+\d^- +\d^+\d^+) 			
\ee

When $G_{mn}^{(2)}(x_1,x_2)$  is properly coupled to the cusp contour, the derivatives in $\dot x_1^m \dot x_2^n D_{mn}$ read

\be
\dot x_1^+ \dot x_2^+ (\d^2 + \d^T\d^T) + \dot x_1^T \dot x_2^T \d^+\d^+ - (\dot x_1^+ \dot x_2^T + \dot x_1^T \dot x_2^+) \d^+\d^T
\ee

which, using the fact that $\dot x_1 \cdot \dot x_2= \dot x_1^+ \dot x_2^- + \dot x_1^- \dot x_2^+  -\dot x_1^T \dot x_2^T$, can be written in the more appealing form

\be
\label{eq:derivatives-2}
\begin{split}
 \dot x_1^m \dot x_2^n D_{mn} =& \\
=&- \dot x_1 \cdot \dot x_2 \d^+\d^+ +\dot x_1^+\d^+ (\dot x_2^- \d_- + \dot x_2^+ \d_+ + \dot x_2^T \d_T) + \dot x_2^+\d^+ (\dot x_1^- \d_- + \dot x_1^+ \d_+ + \dot x_1^T \d_T)\\
=& - \dot x_1 \cdot \dot x_2 \d^+\d^+ +\frac{d}{ds_1}\dot x_2^+\d^+ - \frac{d}{ds_2} \dot x_1^+\d^+\\
\end{split}
\ee

Supplying the equation above with the results (\ref{eq:derivatives-2}) and following, one gets

\be
\label{eq:twoloopint1-2}
\dot x_1^m \dot x_2^n G_{mn}^{(2)}(x_1,x_2)=
 \frac{MN^2}{k^2}\left[ -\frac{\dot x_1 \cdot \dot x_2 }{(x_1-x_2)^2} + \half\left( \dot x_2^+ \frac{d}{ds_1} - \dot x_1^+ \frac{d}{ds_2} \right) \frac{1}{[x^+]}\log\left( -\frac{x^2}{(x^T)^2} \right) \right]
\ee

Note that the first summand above is, up to a total derivative, nothing but the one-loop correction to the CS propagator in covariant (Feynman) gauge computed in \cite{Drukker:2008zx}. In the former gauge one has additional contributions too, which are half-total derivatives, and can be interpreted as correlation functions of a single leg with an external operator. Let us indicate respectively with $L^{(2)}_{\rm vector}, L^{(2)}_{\rm ext-1}$ and $L^{(2)}_{\rm ext-2}$ the three contributions to the integral of (\ref{eq:G2}). Firstly we compute

\be
L^{(2)}_{\rm vector} = -\int_{-L}^{-\gep} \dd s_1\, \int_\gep^L \dd s_2\,  \frac{\dot x_1 \cdot \dot x_2 }{(x_1-x_2)^2} = -\int_{\gep}^{L} \dd s_1\, \int_\gep^L \dd s_2\,\frac{\dot x_1 \cdot \dot x_2 }{2 \dot x_1 \cdot \dot x_2 s_1 s_2}= -\half\left( \log\frac{L}{\gep} \right)^2
\ee 

Then we consider the semi-total derivative pieces

\be
\begin{split}
L^{(2)}_{\rm ext-1} =& \half \int_{-L}^{-\gep} \dd s_1\, \int_\gep^L \dd s_2\, \dot x_2^+ \frac{d}{ds_1} \frac{1}{[x^+]}\log\left( -\frac{x^2}{(x^T)^2} \right)\\
=& \frac{1}{4} \left( \log\frac{L}{\gep} \right)^2 - \half \log\frac{L}{\gep} \log\frac{(-2 \dot x_1 \cdot \dot x_2)}{(\dot x_2^T)^2} + {\rm finite}
\end{split}
\ee

and 

\be
\begin{split}
L^{(2)}_{\rm ext-2} =& -\half \int_{-L}^{-\gep} \dd s_1\, \int_\gep^L \dd s_2\, \dot x_1^+ \frac{d}{ds_2} \frac{1}{[x^+]}\log\left( -\frac{x^2}{(x^T)^2} \right)\\
=& \frac{1}{4} \left( \log\frac{L}{\gep} \right)^2 - \half \log\frac{L}{\gep} \log\frac{(-2 \dot x_1 \cdot \dot x_2)}{(\dot x_1^T)^2} + {\rm finite}
\end{split}
\ee

Note that pure divergences in the bubble graph cancel out between the three terms above. At the end we must consider single-leg contributions, that quite unexpectedly do not vanish, as we already remarked. These terms need a careful treatment of contact divergences, {\it i.e.} divergences of the propagator arising when its two ends collide against each other. To this end, note that the Mandelstam-Leibbrandt prescription $[x^+]=x^+ +\eta\,{\rm sign}(x^-)$ on the spurious poles at $x^+=0$ acts like a framing on the contour, shifting in facts by a vector $\eta$ the integration of one of the $s$'s to a "frame contour" $\cC'$ infinitely close to $\cC$. Bearing this in mind, and that the framing vector must be orthogonal to the contour itself $\eta\cdot \dot x_i=0$, we can safely remove the $\gep$ cutoff near the origin $s=0$. Then, considering that $\dot x_i^2=0$ in the present case, the first term in $D_{mn}$ in equation (\ref{eq:derivatives-2}) drops, and (\ref{eq:twoloopint1-2}) becomes 

\be
\label{eq:singleleg}
\half\left( \frac{d}{ds_1} - \frac{d}{ds_2} \right) \frac{\dot x_i^+}{[\dot x_i^+(s_1-s_2)]}\log\left( -\frac{[\dot x_i(s_1-s_2)]^2}{[\dot x_i^T(s_1-s_2)]^2} \right)
\ee

Using the symmetries of this integrand it is easy to show that the two single-leg diagrams merge together, and the two total derivatives contribute equally, hence the total contribution of these diagrams is 

\be
\begin{split}
& \int_{0}^{L} \dd s_1\, \int_{0}^{L} \dd s_2\, \half\left( \frac{d}{ds_1} - \frac{d}{ds_2} \right) \frac{1}{s_1-s_2+\eta}\log\left( -\frac{x_1^2}{(x_1^T)^2} \right)\\
=& -\left(\log\frac{L}{\eta} \right)^2 - \log\frac{L}{\eta}\,\log (x_1^T)^2 + \cO\left( \frac{\eta}{L} \right)
\end{split}
\ee

See equation (\ref{eq:singleleg}) and following for more details. Note that one can effectively trade $\eta$ for $\gep$, indeed had we removed the cutoff from the very beginning in all integrals and used the framing instead, we would have ended up with same results, up to the identification $\eta\to\gep$. This is also clear from the fact that the presence of $\gep$ is only needed to regularise the contact divergence in $s=0$, which is the same job that $\eta$ does. 
Finally, putting all the two-loops contributions together and considering the contribution of the $\hat A$ field we then find

\be
\label{eq:cusp-two-loops}
\left< W^{(2)}(\cC) \right> =\left( \frac{NM}{k^2} \right) \left\{ -\frac{1}{2}\left( \log\frac{L}{\gep} \right)^2 - \frac{1}{2} \log\frac{L}{\gep} \log\frac{(-2 \dot x_1 \cdot \dot x_2)}{(\dot x_1^T \dot x_2^T)} + {\rm finite} \right\}
\ee

which can be compared with the result obtained in Landau gauge in \cite{Henn:2010ps} setting $N=M$ and expanding in the limit where the dimensional regularization parameter $\gep_D\to 0$

\begin{multline}
W^{(2)}_{\rm Dim}=\left( \frac{N}{k} \right)^2\left[-\frac{1}{8} \frac{(-(2 \dot x_1 \cdot \dot x_2) \mu^2)^{2\gep_D}}{\gep_D^2} \right] = \\ =\left( \frac{N}{k} \right)^2 \left[ -\frac{1}{8\gep_D^2}- \frac{1}{4\gep_D} \log(-(2 \dot x_1 \cdot \dot x_2) \mu^2) + {\rm finite} \right]
\end{multline}

The two results agree up to the identification of the regularisation parameters
$\frac{1}{2\gep_D} \to \log\frac{L}{\gep}$ and regularizations scales $\mu^2 \to \dot x_1^T \dot x_2^T$. A fact that could have been expected, but that actually finds a rather non-trivial verification in the two expressions above. Moreover, we want to stress that single-leg diagrams are crucial for recovering the correct result at the perturbative level.


\subsection{The deformed case}

Let us now turn our attention to the deformed contour of (\ref{eq:cuspdeformed}). Computing (\ref{eq:onegluon}) and (\ref{eq:twogluons}) above we have been a little sketchy in handling  $\Theta$-function constraints. An alternative way of proceeding is to regularise the $\gd$-function in the propagator in the following way

\be
\label{eq:CSpropreg}
G^{(1)}_\eta(s_1,s_2) = - \lim_{\eta\to 0}\frac{1}{k \pi} \frac{1}{[x^+(s_1) -x^+(s_2)]}  \frac{\eta}{\eta^2 + [x^T(s_1) -x^T(s_2)]^2} 
\ee

and computing the expectation value of a single CS exchange on the deformed contour. 
Using the contour (\ref{eq:aphimap}) and the map (\ref{eq:cuspdeformed}), the regularised propagator above takes the form

\be
\begin{split}
L_A^\eta =& -\lim_{\eta\to 0} \frac{\eta N^2}{k\pi} \int_{-L}^{-\gep} ds_1\, \int_\gep^L ds_2\, \left(2\e^{-\phi}\ga^{-1} - 2\e^{-\phi}\ga +\ga\gb \right)\\
& \times \left[ s_1(\gb-1)-s_2(\gb+1) \right]^{-1} \left[ \eta^2 + \left( \e^{-\phi}(s_1-s_2)\frac{\gb^2}{\ga} + (s_1+s_2)\ga \right)^2 \right]^{-1}\\
=&-\lim_{\eta\to 0} \frac{\eta N^2}{k\pi} \int_\gep^L ds\, \int_{\gep/s}^{L/s} dx\, \left(2\e^{-\phi}\ga^{-1} - 2\e^{-\phi}\ga +\ga\gb \right) \left[ (\gb-1)-x(\gb+1) \right]^{-1} \\
&\times \left[\eta^2 +s^2 \left( \left(\e^{-\phi} \frac{\gb^2}{\ga} -\ga \right)x + \left(\e^{-\phi} \frac{\gb^2}{\ga} +\ga \right)  \right)^2  \right]^{-1}
\end{split}
\ee

where we have introduced the short-hands $\ga=\sin\gth,\,\gb=\cos\gth$ and changed variables to $s_1\to -s,\,s_2\to xs$. In the limit where $L\to\infty,\,\gep\to 0$, the integral above evaluates to

\be
\label{eq:LAreg}
\left< W_a^{(1)} \right> = \frac{N}{k}  \left[ \log\frac{L}{\gep} - 2 \phi \left( \e^{-\phi}\gb/\ga^2 +\frac{1}{\ga} \right) \right] 
\ee
	
Quite interestingly the regularized propagator can capture subleading corrections to the expectation value of the deformed Wilson loop $W(\cC_{\rm def})$ which, despite the seemingness of the deformation of (\ref{eq:aphimap}), do make sense. Indeed, it was claimed in \cite{Griguolo:2012iq} that the v.e.v of the generalized cusp anomalous dimension computed in Euclidean spacetime, when analytically continued to Lorentz signature,  at one-loop order leads to

\be
\label{eq:one-fermion}
\lim_{\phi\to\ii\infty} \Gamma_{gen}(\phi)= \half\frac{N}{k}(\mu L)^{2\gep_D} \left(\frac{1}{\gep_D} - \e^{-\frac{\phi}{2}}\phi \right) + \cO(N^2/k^2)
\ee

where $2\gep_D=3-D$ is the dimensional regulator and $\mu$ a mass scale. Note that the two result are strikingly similar. The leading contributions for $L$ large coincide up to the identification $\frac{1}{2\gep_D} = \log\frac{L}{\gep}$, and the subleading corrections in the cusp angle $\phi$ display the same damped structure. From this rather nontrivial result it is natural to expect that the leading contribution to the cusp anomalous dimension cannot come from ladders of gauge field propagators, as their $\phi$ dependence is exponentially suppressed and, most of all, independent of the any UV/IR scale.

\vspace{10mm}
{\bf Two-loops diagrams}\\

Diagrams contributing at order $1/k^2$ are the same as in the strictly light-like case. The double exchange diagram will very likely exponentiate the one loop result, and hence will not contribute to the leading order of cusp anomalous dimension due to the heavily suppressed nature of its $\phi$ dependence. So, consider again the one-loop corrected propagator (\ref{eq:twoloopint1-2}). It generates three main contributions, an exchange one and two single-leg pieces as before. The computation of these diagrams is analogous to the previous case, up to the fact that one must consider one more regulator in this case : $a\sim|\dot x_i|$. The details of this computation are in Appendix \ref{sec:on-deformed}. The exchange contribution amounts to (up to a group factor)

\be
\begin{split}
W_{\rm exch}^{(2)}(\cC_{\rm def}) =& \int_{-L}^{-\gep} \dd s_1\, \int_\gep^L \dd s_2\, \left[ -\frac{\dot x_1 \cdot \dot x_2}{ 2a\ga (s_1^2-s_2^2) -2(\dot x_1 \cdot \dot x_2)s_1 s_2} \right. \\
&\left. + \half\left( \dot x_2^+ \frac{d}{ds_1}- \dot x_1^+ \frac{d}{ds_2}\right) \frac{1}{\dot x_1^+ s_1-\dot x_2^+ s_2}\log\left( -\frac{2a\ga (s_1^2-s_2^2) -2(\dot x_1 \cdot \dot x_2)s_1 s_2}{[\ga(s_1-s_2)-a(s_1+s_2)]^2} \right) \right]\\
=&-\half \log\frac{L}{\gep}\,\log(a) + \cO\left( \frac{\gep}{L} \right)
\end{split}
\ee

For what concerns single-leg pieces, one has to frame the integration contour as was done before in (\ref{eq:singleleg}) using a "small" vector $|\eta|\sim 0$, $\eta\cdot \dot x_i=0$. Due to the symmetries of the problem they can be merged into a single global integral just as in the light-like case

\be
\begin{split}
W_{\rm ext}^{(2)}(\cC_{\rm def}) =& 2\int_0^L \dd s_1\, \int_0^L \dd s_2\,\left[- \frac{2a\ga}{2a\ga(s_1-s_2)^2 + \eta^2} \right. \\
& +\left.\half \left( \frac{d}{ds_1}-\frac{d}{ds_2} \right) \frac{1}{s_1-s_2+\eta} \log\left( \frac{2a\ga(s_1-s_2)^2+\eta^2}{[(\ga-a)(s_1-s_2)]^2}  \right) \right]\\
=& 2\log\frac{L}{\eta}\, \left(\log(a)+1\right) + \cO\left( \frac{\gep}{L} \right)
\end{split}
\ee

Accounting for a  totally equivalent contribution coming from the second gauge group $\hat A$, we get to the following expression for the two-loop expectation value of the deformed Wilson loop

\be
\left<W_a^{(2)} \right>= 1 + \frac{N-M}{k}  \left[ \log\frac{L}{\gep} - 2 \phi \left( \e^{-\phi}\gb/\ga^2 +\frac{1}{\ga} \right) \right] +  \half \frac{NM}{k^2} \log\frac{L}{\gep}\, \left(\log(a)+1\right)
\ee

which for large values of the velocity (cusp angle) $\phi$, corresponding to small $a$, and setting $N=M$ becomes

\be
\label{eq:wcad}
W^{(2)}(\phi\to\infty)= 1 - \frac{\phi}{2}\gl^2  \log\frac{L}{\gep}
\ee

that correctly reproduces the  universal cusp anomalous dimension at two-loops. One more comment is in order here. Substituting the map (\ref{eq:aphimap}) into the $\gl^2$ term of $W^{(2)}(\cC_{\rm def})$ and keeping subleading (in $\phi$) terms one has

\be
-\frac{\gl^2}{2} \log\frac{L}{\gep}\left( \phi + \log \dot x_1 \cdot \dot x_2 \right)
\ee

which nothing is but the strictly light-like result (\ref{eq:cusp-two-loops}) up to the identification $\phi \leftrightarrow \log\frac{L}{\gep}$ for $\phi\to\infty,\,L\to\infty,\,\gep\to 0$. So, although it might seem counter intuitive at first glance, the cusp anomalous dimension is sensitive to the global light-like structure of the Wilson loop, rather than the local geometry of the cusp point, in that the velocity $\phi$ itself becomes a global parameter.


\section{Non-perturbative analysis}
\label{sec:nonpert}

Motivated by the one and two-loops results, we now address the problem of determining the asymptotic behaviour of the Wilson loop coupled to the deformed contour $\cC_{\rm def}$ at large values of the 't~Hooft coupling $\gl=\frac{N}{k}$. From the one and two-loops results (\ref{eq:LAreg}) and (\ref{eq:wcad}) for CS propagators,  it is legitimate to expect an exponentiation of the one-loop result when resumming planar ladders. Also we might conjecture, and comfortably, that resummation of free propagators will not give any contribution to the cusp anomalous dimension, which is usually understood as the $\phi$ coefficient in the large $\phi$ expansion of the light-like cusp. Moreover, from the two-loops expression (\ref{eq:wcad}), it is rather tempting to conjecture that, unlike the four dimensional case \cite{Makeenko:2006ds}, ladders of corrected propagators alone could reproduce the four-dimensional-like $\sqrt{\gl}\phi$ behaviour at strong coupling \cite{Forini:2012bb}. Such a feeling arises mainly from the observation that the gauge field three-vertex plays no role for the evaluation of this observable.

\begin{wrapfigure}[14]{l}{40mm}
\includegraphics[width=.2\textwidth]{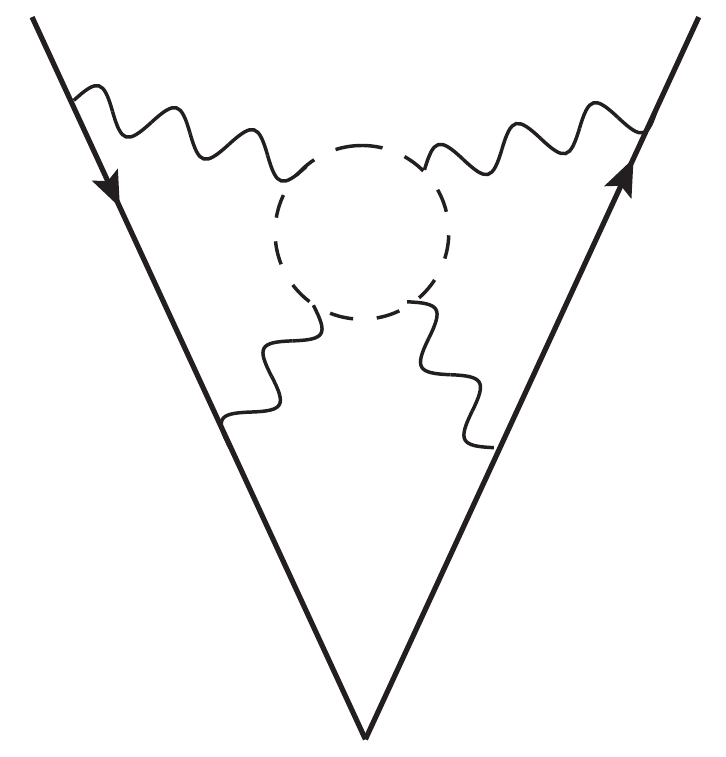}
\caption{\label{fig:highervertex} Although the gauge field three-vertex is gauged away, higher order $n-$points interactions arise as effective vertices through loops of fermionic and scalar matter.}
\end{wrapfigure}

As a matter of fact it is ruled out of the action by the gauge choice from the very beginning. Hence gauge fields interact only through higher order graphs like the one in Figure \ref{fig:highervertex}. On the other hand, the deformation (\ref{eq:cuspdeformed}) between the locally $\half$ BPS Wilson loop operator of \cite{Griguolo:2012iq} and the light-like contour might not be a smooth map at strong coupling, since higher loop fermionic and scalar diagrams may not be sufficiently suppressed, hence our observable is not bound to be dual to the semiclassical string solution.

\subsection{Resumming planar ladders}
\label{sec:planladd}

The main goal of this section is to resum ladders of free CS propagators. The Bethe-Salpeter equation for planar ladder diagrams on the deformed contour $\cC_{\rm def}$ reads

\be
\label{eq:BS-ladder}
B(L_1,L_2,\gep_1,\gep_2) = 1 +\int_{-L_1}^{-\gep_1} ds_1\, \int_{\gep_2}^{L_2} ds_2\, B(s_1,s_2,\gep_1,\gep_2) \dot x_1^m \dot x_2^n G^{(1)}_{mn}(s_1,s_2,a)
\ee

where the kernel $G^{(1)}(s_1,s_2,a)$ is the regularised propagator of (\ref{eq:CSpropreg}). Again we use the map

\be
a = \e^{-\gth}\frac{\sin^2\phi -1}{\sin\phi}
\ee

between the deformation parameter $a$, the (infinite) cusp angle $\phi$ and the (finite) geometric angle $\gth$. Note that the leading asymptotic behaviour of $B(L_1,L_2,\gep_1,\gep_2)$ for large values of $\phi$ should not depend on the geometric angle $\gth$ as the $\phi\sim\infty$ expansion is a $\log\,a$ expansion at leading order. Changing sign of $s_1$ and then differentiating (\ref{eq:BS-ladder}) with respect to $L_1$ and $L_2$ produces the equation

\be
\label{BS-ladder-differential}
\frac{d}{dL_1}\frac{d}{dL_2} B(L_1,L_2,\gep_1,\gep_2) - \dot x_1^m \dot x_2^n G^{(1)}_{mn}(-L_1,L_2,a)B(L_1,L_2,\gep_1,\gep_2) =0
\ee

which is dominated by the region where both $L_1,L_2$ are large and $L_1\sim L_2$. From the integral equation above it is easy to read the boundary conditions for Bethe-Salpeter solution 

\be
 B(\gep_1,L_2,\gep_1,\gep_2)= B(L_1,\gep_2,\gep_1,\gep_2) =1
\ee

 Making avail of the extra degree of freedom $\gth$ one can get to a simplified differential equation for $B$. Namely, we can substitute in the equation above the complete kernel

\be
\begin{split}
K(s,t,\phi,\gth)=& \dot x_1^m \dot x_2^n G^{(1)}_{mn}(-L_1,L_2,a)\\
=&\gl \frac{2\eta}{\pi} \frac{\e^{-\phi}\gb^2 +\ga\gb}{\ga\left[ (s-t) - \gb(s+t) \right] \left[ \eta^2 + \left( \e^{-\phi}(s+t)\frac{\gb^2}{\ga} +(s-t)\ga \right)^2 \right] }
\end{split}
\ee

where $\ga=\sin\gth,\, \gb=\cos\gth$ and $\gl=\frac{N}{k}$, with its value for some choice of $\ga,\,\gb$. To this end note that

\be
\begin{split}
& \eta^2 + \left( -\e^{-\gth}(L_1+L_2)\frac{\gb^2}{\ga} - (L_1-L_2)\ga \right)^2\\
=& \eta^2  - 2 L_1 L_2 \left(\ga^2 -\e^{-2\gth}\frac{\gb^4}{\ga^2}\right) + (L_1^2 + L_2^2) \left( \ga^2 + \e^{-2\gth}\frac{\gb^4}{\ga^2} \right)\\
& + 2\e^{-\gth} \gb^2 (L_1+L_2) (L_1-L_2)
\end{split}
\ee

and hence, setting 

\be
\ga^4 + \e^{-2\gth} \gb^4 = 0
\ee

the equation above gets simplified to 

\be
\eta^2  - L_1 L_2 \e^{-2\gth}\frac{\gb^4}{\ga^2}\sim \left(\eta + 2\ii L_1 \e^{-\gth}\frac{\gb^2}{\ga} \right) \left(\eta -\ii 2 L_2 \e^{-\gth}\frac{\gb^2}{\ga} \right)
\ee

up to terms of order $\cO(L_1-L2)<<L_1 + L_2$ which we expect to be subleading in the large $L_1\sim L_2$ limit, and one has a simplified version of the kernel $K$

\be
\label{eq:siply-kernel}
\hat K(L_1,L_2) = \hat K_1(L_1) \hat K_2(L_2)= \gl \frac{\eta}{\pi} \frac{\e^{-\phi}\frac{\gb}{\ga} +1}{L_1 \left(\eta + 2\ii L_1 \e^{-\gth}\frac{\gb^2}{\ga} \right) \left(\eta -\ii 2 L_2 \e^{-\gth}\frac{\gb^2}{\ga} \right) }
\ee

Thanks to this simple observation, (\ref{BS-ladder-differential}) becomes a   separate variables differential equation, that can be decomposed in the two first order equations

\be
\label{eq:BS-ladder-separated}
\begin{split}
&\frac{d}{dL_1}B_1(L_1,\gep_1) - \frac{ \eta \sqrt{\gl}}{L_1 \left(\eta + 2\ii L_1 \e^{-\gth}\frac{\gb^2}{\ga}\right)}B_1(L_1,\gep_1) =0 \\
&\frac{d}{dL_2}B_2(L_2,\gep_2) - \frac{\sqrt{\gl}\left(\eta - 2\ii L_2 \e^{-\gth}\frac{\gb^2}{\ga} \right)}{\left(\eta - 2\ii L_2 \e^{-\gth}\frac{\gb^2}{\ga}\right)}B_2(L_2,\gep_2) =0
\end{split} 
\ee

supported by the boundary conditions 

\be
B_1(\gep_1,\gep_1)=B_2(\gep_2,\gep_2) = 1
\ee

and the requirement of $L_1 \leftrightarrow L_2$ symmetry. Differentiating (\ref{eq:BS-ladder}) with respect to $\gep_1,\,\gep_2$ and proceeding along the same line, one can write equations analogous to (\ref{eq:BS-ladder-separated}) in these two variables. Solving both and combining the results in the limit where $L_1=L_2=L\to\infty$ and $\gep_1=\gep_2=\gep\to 0$ one has the asymptotic behaviour of the Bethe-Salpeter solution for planar ladders

\be
\label{eq:ladders-asymp}
B(L,\gep,\phi) = \e^{ \frac{\sqrt{\gl}}{\gb} \left( \log\frac{L}{\gep} + (\gb-1)\phi \right) + \cO(\ga,\gb)  } 
\ee

This result is even clearer at large cusp angle $\phi\to \infty$, recalling that $\gb^2=1-\ga^2$ 

\be
\lim_{\phi\to\infty} B(L,\gep,\phi) = \exp\left\{ \sqrt{\gl} \left( \log\frac{L}{\gep} + \e^{-\phi}\phi \right) + \cO(\ga,\gb)  \right\} 
\ee

which, up to possible $\cO(\ga,\gb)$ subleading contributions, exponentiates the one loop result of \ref{eq:LAreg}. Furthermore, we expect that subleading correction must be of the form $\sim \log\frac{L_1}{L_2}$ in the limit where $L_1=L_2=L\to\infty$ to make our derivation self-consistent.  It is straightforward to compute them, indeed let us introduce the new variables

\be
\tau=L_1+L_2, \qquad \gg=\frac{L_1-L_2}{\tau}
\ee

in terms of which equation (\ref{BS-ladder-differential}) becomes

\be
\left(\d^2_\tau - \frac{1}{\tau^2}\d^2_\gg \right)B(\tau,\gg) = K(\tau,\gg)B(\tau,\gg)
\ee

Using the large $\tau$ asymptotics (\ref{eq:ladders-asymp}) and the ansatz 

\be
B(\tau,\gg)=B(\tau)\Gg(\gg)= B(\tau) \exp\left(\sqrt{\gl}\sum_n \gl^{-\frac{n}{2}} f_n(\gg) \right)
\ee

the latter becomes a second order differential equation in $\gg$, that at leading order in $\sqrt{\gl}$ reads

\be
(\d_\gg f_0(\gg))^2 = 1 - \tau K(\tau,\gg)
= 1-  \frac{2\eta}{\pi} \frac{2 \gb \tau}{(\gg - \gb) \left[ \eta^2 + \tau^2 \e^{-\phi} \gg \right]} 
\ee

Taking the square root on the r.h.s. and integrating, the solution for $f_0$ can be written in closed form in terms of an incomplete elliptic integral of the second kind. Taking the $\eta\to 0$ limit one has

\be
f_0= \half \gb\, \mathbb{E}\left( {\rm Arcsin}\left(\frac{2\gg-\gb}{\gb}\right)\bigg| 1 \right) = \gg-\half \gb + \cO(\gg^2)
\ee

Note that for $L_1\sim L_2$ large, the latter is equivalent to 

\be
\log\frac{L_1}{L_2} = \log\left( 1- \frac{L_1-L_2}{L_2} \right) \sim \log\left(1-\half\gg\right)  = -\half\gg + \cO(\gg^2)
\ee 

and hence, as expected.
The strong coupling analysis of ladders of free gauge propagators emphasises three noticeable results. Firstly,  for ABJ theories,  the Bethe-Salpeter kernel in the ladder approximation at large value of the 't~Hooft coupling $\gl$  exponentiates the one loop result at leading order in the UV/IR regularisation parameters $\log\frac{L}{\gep}$. Secondly, it does not contribute to the cusp anomalous dimension (usually understood as the coefficient of $\frac{\phi}{2}$) because of the fact that, at large cusp angle $\phi$, the $\phi$ dependence is insensitive of the regularization scale $\log\frac{L}{\gep}$. Thirdly, as was already pointed out in \cite{Griguolo:2012iq}, the contribution of (\ref{eq:ladders-asymp}) can be interpreted as an open contour renormalisation factor since it does not depend on any of the kinematical variables, not only the cusp angle.


\subsection{Ladders of corrected propagators}
\label{sec:corrladd}

The corrected CS propagator captures at two-loops the correct cusp anomalous dimension of SCS theories and to obtain the right result it turns out to be crucial to include single-leg diagrams in the computation. Therefore we will consider the following Bethe-Salpeter kernel expanded at order $\gl^2$  

\be
\label{eq:BSK-twoloops}
K^{(2)}(s,t,s',t')= \gl K_0(s,t)\gd(s-s')\gd(t-t') + \gl^2 \left[ K_1(s,t)\gd(s-s')\gd(t-t')+K_2(s,t,s',t')  \right]
\ee

where

\be
\begin{split}
K_0(s,t) =&  \frac{2\eta}{\pi} \frac{\e^{-\phi}\gb^2 +\ga\gb}{\ga\left[ (s-t) - \gb(s+t) \right] \left[ \eta^2 + \left( \e^{-\phi}(s+t)\frac{\gb^2}{\ga} +(s-t)\ga \right)^2 \right] }\\
K_1(s,t) =&  -\frac{\dot x_1 \cdot \dot x_2 }{(x_1(s)-x_2(t))^2} \\
&+ \half\left( \dot x_2^+ \frac{d}{ds} - \dot x_1^+ \frac{d}{dt} \right) \frac{1}{[x_1^+(s)-x_2^+(t)]}\log\left( -\frac{(x_1(s)-x_2(t))^2}{(x_1^T(s)-x_2^T(t))^2} \right) \\
K_2(s,t) =& -\frac{\dot x_1^2 }{(x_1(s)-x_1(s'))^2} -\frac{\dot x_1^2 }{(x_2(t)-x_2(t'))^2} \\
&+ \half \dot x_1^+\left( \frac{d}{ds} -  \frac{d}{ds'} \right) \frac{1}{[x_1^+(s-s')]}\log\left( -\frac{(x_1(s)-x_1(s'))^2}{(x_1^T(s)-x_1^T(s'))^2} \right) \\
&+ \half \dot x_2^+\left( \frac{d}{dt} -  \frac{d}{dt'} \right) \frac{1}{[x_2^+(t-t')]}\log\left( -\frac{(x_2(t)-x_2(t'))^2}{(x_2^T(t)-x_2^T(t'))^2} \right) \\
\end{split}
\ee

which in pictorial form reads

\be
\raisebox{-.05\textwidth}{\includegraphics[width=.12\textwidth]{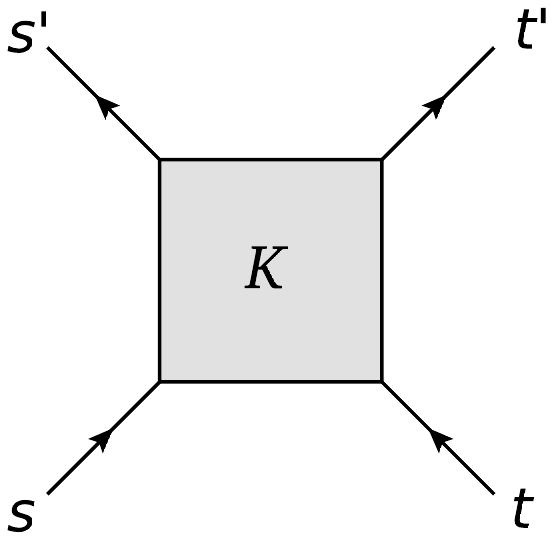}} =
\gl\,\, \raisebox{-.05\textwidth}{\includegraphics[width=.12\textwidth]{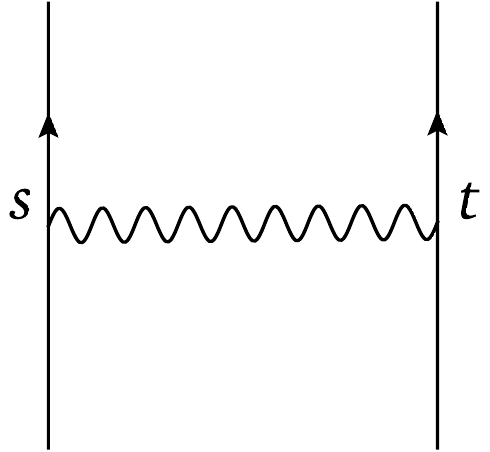}}+
\gl^2 \, \left[ \, \raisebox{-.05\textwidth}{\includegraphics[width=.12\textwidth]{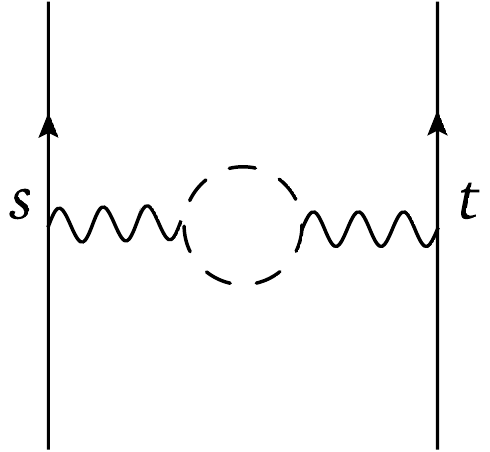}} \, + \,
\raisebox{-.05\textwidth}{\includegraphics[width=.12\textwidth]{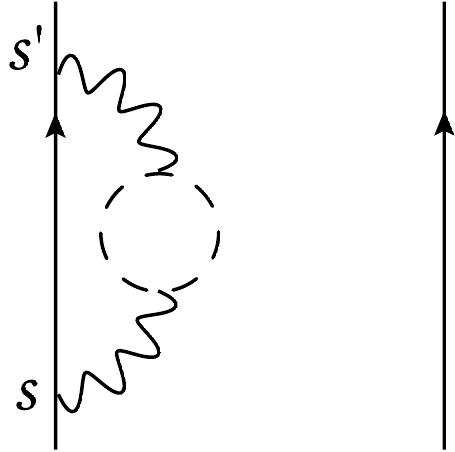}} \, + \, 
\raisebox{-.05\textwidth}{\includegraphics[width=.12\textwidth]{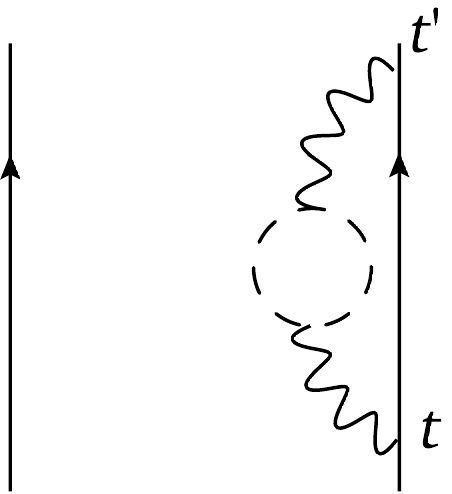}} \,
\right]
\ee

After reversing the sign of $s,\,s'$, the proper Bethe-Salpeter equation reads 

\be
G(S,T)=1+ \int_\gep^S ds\,\int_\gep^T dt\, \int_s^S ds'\, \int_t^T dt'\, K^{(2)}(s,t,s',t')\, G(s,t)
\ee

with boundary conditions $G(\gep,T)=G(S,\gep)=1$. Differentiating w.r.t. $S$ and $T$ one has

\be
\label{eq:BS-twoloops}
\frac{\d}{\d S}\frac{\d}{\d T}G(S,T) = \int_\gep^S ds\,\int_\gep^T dt\, K^{(2)}(s,t,S,T) G(s,t)
\ee

which is a quite complicated integro-differential equation, however we are only interested in solving this equation in the limit where $S$ and $T$ are both very large, so we define new variables

\be
\tau=S+T \to\infty, \qquad \gs=S-T\to 0, \qquad \frac{\d}{\d S}\frac{\d}{\d T}=\frac{\d^2}{\d\tau^2}-\frac{\d^2}{\d\gs^2}
\ee

It is convenient to separate the contribution of tree-level propagators computed in equation (\ref{eq:ladders-asymp}) in  the previous section 

\be
G(\tau,\gs)= G^{(0)}(\tau,\gs)\, G^{(1)}(\tau,\gs)
\ee

Substituting it into (\ref{eq:BS-twoloops}) the contribution of the two  derivatives acting on $G^{(0)}$ cancels that of $K^{(0)}$ and we are left with

\begin{multline}
\label{eq:BS-difform}
\left( \d^2_\tau -\d^2_\gs \right) G^{(1)}(\tau,\gs) + \frac{2\sqrt{\gl}}{\tau} \left( \d_\tau-\d_\gs \right) G^{(1)}(\tau,\gs) =
\frac{\gl^2}{G^{(0)}(\tau,\gs)} \int_\gep^{\frac{\tau+\gs}{2}} du\, \int_\gep^{\frac{\tau-\gs}{2}} dv\, \\ G(\tau-u-v,\gs-u+v) \left[ K_1(\tau-u-v,\gs-u+v)\gd(u)\gd(v)+K_2(u,v)  \right]
\end{multline}

while the boundary conditions now read $G(\tau=\pm \gs,\gs)=1$.
On the right-hand side of the equation above compares the full Bethe-Salpeter wave function $G=G^{(0)}G^{(1)}$. The first contribution to the kernel has point-like insertions, so the integral is trivial and $G^{(0)}$ can be factored out

\be
\begin{split}
&\frac{\gl^2}{G^{(0)}(\tau,\gs)} \int_\gep^{\frac{\tau+\gs}{2}} du\, \int_\gep^{\frac{\tau-\gs}{2}} dv\, G(\tau-u-v,\gs-u+v)\, K_1(\tau-u-v,\gs-u+v)\gd(u)\gd(v)\\
=&\gl^2 G^{(1)}(\tau,\gs) K_1(\tau,\gs)
\end{split}
\ee

while the second can be consistently simplified in the $\tau\to\infty,\,\gs\to 0$ limit, where the kernels become

\be
\begin{split}
K_1(\tau) =& -\frac{\gb^2}{\ga^2}\frac{\log a}{\tau^2}\\
K_2(u) =& \frac{d}{du} \frac{1}{u+\eta}\log a
\end{split}
\ee

It is clear that in this limit the integrals receive contribution only in the $u\sim v\sim 0$ region, moreover the two single-leg terms are symmetric in the exchange of $u$ and $v$, so 

\be
\begin{split}
&\gl^2 \e^{-\sqrt{\gl}\log\tau} \int_\gep^{\frac{\tau}{2}} du\, \int_\gep^{\frac{\tau}{2}} dv\, \e^{\sqrt{\gl}\log(\tau-u-v)}G^{(1)}(\tau-u-v)K_2(u,v) \\
=&\gl^2 \int_\gep^{\frac{\tau}{2}} du\, \int_\gep^{\frac{\tau}{2}} dv\, \e^{-\sqrt{\gl}\frac{u+v}{\tau}}G^{(1)}(\tau-u-v)\frac{d}{du}\frac{1}{u+\eta}\log a
\end{split}
\ee

We attempt to solve the resulting equations by means of an exponential ansatz for the BS wave function $G^{(1)}(\tau)=\e^{F(\tau)}$, then by differentiation of (\ref{eq:BS-difform}) we obtain

\be
\begin{split}
&F''+(F')^2 + \frac{2\sqrt{\gl}}{\tau}\,F' = \gl^2 K_1 + 4\gl^2 \int_\gep^{\frac{\tau}{2}} du\, \int_\gep^{\frac{\tau}{2}} dv\, \e^{-\sqrt{\gl}\frac{u+v}{\tau}} \e^{-(u+v)F'(\tau)} \frac{d}{du}\frac{1}{u+\eta}\log a
\end{split}
\ee

where we have used the fact that $F(\tau-u-v)-F(\tau)=-(u+v)F'(\tau)+\cO((u+v)^2)$ for small $u,\,v$. Integrating by parts over $u$ and $v$ and expanding for large $\gl$, one gets the following differential equation for $F(\tau)$

\be
\label{eq:BS-F}
F''+(F')^2 + \frac{2\sqrt{\gl}}{\tau}\,F'-\gl \tau^2 \phi - \gl^2\frac{\phi \gb^2}{\tau^2\ga^2}=0
\ee

which integrates to \footnote{We have omitted subleading terms in order to find an analytical result.}

\be
\begin{split}
F(\tau)= &\cC_1\left\{\half \sqrt{\gl}\sqrt{1+\tau^4\phi+\gl\phi\cot^2\gth} + \sqrt{\gl}\log\tau(-1 +\sqrt{1+\gl\phi\cot^2\gth}) \right. \\
&\left.- \half \sqrt{\gl} \sqrt{1+\gl\phi\cot^2\gth}\, \log\left[1+\gl\phi\cot^2\gth +\sqrt{1+\gl\phi\cot^2\gth}\sqrt{1+\tau^4\phi+\gl\phi\cot^2\gth} \right] \right\}\\
&+\cC_2\left\{-\half \sqrt{\gl}\sqrt{1+\tau^4\phi+\gl\phi\cot^2\gth} + \sqrt{\gl}\log\tau(-1 -\sqrt{1+\gl\phi\cot^2\gth}) \right. \\
&\left.+ \half \sqrt{\gl} \sqrt{1+\gl\phi\cot^2\gth}\, \log\left[1+\gl\phi\cot^2\gth +\sqrt{1+\gl\phi\cot^2\gth}\sqrt{1+\tau^4\phi+\gl\phi\cot^2\gth} \right] \right\}\\
\end{split}
\ee

with integration constants $\cC_1,\,\cC_2$ (we have traded $\tau/\gep$ for $\tau$ for simplicity). Happily enough the $-\sqrt{\gl}\log\tau$ cancels with the corresponding term coming from $G^{(0)}$. Then, we must take the limit in which (\ref{eq:BS-F}) holds, namely for $\frac{\tau}{\gep}>>\sqrt{\gl}$, so we obtain the asymptotic behaviour for the Bethe-Salpter wave function

\be
G(\tau) = \cC_1 \e^{\half\sqrt{\gl\phi}\tau^2} (\gl\phi\cot^2\gth)^{-\half\gl\sqrt{\phi}\cot\gth}
\ee

According to the boundary conditions, $\cC_1$ must be chosen such that $G(\tau)= 1$ when $\tau\sim\gep$, that is easily solved by 

\be
\label{eq:resummcorrladd}
G(\tau) =  \e^{\half\sqrt{\gl\phi}\left(\frac{\tau^2}{\gep^2}-1\right)}
\ee

We want to stress that this result is valid only in the kinematical region where $\frac{\tau}{\gep}>>\sqrt{\gl}$ with $\gl,\tau\to\infty$ and $\gep\to 0$. 

\subsection{Comparison to string theory}
\label{sec:comparison}

Although the result (\ref{eq:resummcorrladd})  displays the $\exp\sqrt{\gl}$ behaviour obtained in string theoretic computations, the $\phi$ dependence is incorrect. To see this, let us consider the expression for the generalised cusp anomalous dimension $\Gg_{\rm cusp}^{AdS}(\gl,\phi,\vartheta)$ in $AdS_4\times\mathbb{C}P^3$ \cite{Forini:2012bb} (which is the same as in $AdS_5\times S^5$ at the classical level \cite{Drukker:2011za} modulo the map $\gl_{SYM}\to 2\pi^2\gl_{CS}$)

\be
\label{eq:gammaads}
\Gg_{\rm cusp}^{AdS}(\gl,\phi,\vartheta)=\sqrt{2\gl} \frac{\sqrt{b^4 + p^2}}{b p} \left[ \frac{(b^2+1)p^2}{b^4+p^2} \mathbb{K}(k^2)-\mathbb{E}(k^2) \right]
\ee

where $\phi,\, \vartheta$ are implicitly determined by the following transcendental equations in terms of the energy $E$ and angular momentum $J$ through $q=-\frac{J}{E}$ and $p=\frac{1}{E}$

\be
\label{eq:adsphitheta}
\begin{split}
\phi =& \pi - \frac{2p^2}{b\sqrt{b^4+p^2}} \left[ \Pi\left(\frac{b^4}{b^4+p^2}\bigg|k^2  \right) - \mathbb{K}(k^2) \right]\\
\vartheta =& \frac{2bq}{\sqrt{b^4+p^2}} \mathbb{K}(k^2)
\end{split}
\ee

being

\be
q^2= \frac{b^2(1-2k^2-k^2 b^2)}{b^2+k^2} \qquad p^2=\frac{b^4(1-k^2)}{b^2+k^2} \qquad 
\ee

We are interested in the $\phi\to\ii\infty,\,\vartheta\to 0$ limit of $\Gg_{\rm cusp}^{AdS}(\gl,\phi,\vartheta)$. This can be achieved by sending the parameter $k^2$ to infinity, so the complete elliptic integral of the first kind goes like

\be
\mathbb{K}(k^2) = \frac{1}{k} \left(-\ii\log k - 2\ii\log2 +\frac{\pi}{2}  \right) + \cO(k^{-2})
\ee
 and the coefficient in front of it in $\vartheta$ behaves as 

\be
\frac{2\ii\sqrt{b^2-2}}{\sqrt{b^2+1}}k
\ee

This is easy sent to zero by choosing $b^2=-2$. The characteristic of the complete integral of the third kind becomes $\frac{b^4}{b^4+p^2}=-k^2$, hence using the relation 

\be
\begin{split}
& \Pi(n|m) = (-n(1-m))(1-n)^{-1} (m-n)^{-1} \Pi(n'|m)) + m(m-n)^{-1} \mathbb{K}(m)\\
& n' = (m-n)(1-n)^{-1} 
\end{split}
\ee

one gets back to the hyperbolic case and has

\be
\begin{split}
\phi =& \pi - \frac{2p^2}{b\sqrt{b^4+p^2}} \left[ -\half\Pi\left(2 | k^2  \right) - \half \mathbb{K}(k^2) \right]\\
=&\pi+ \ii\,2\sqrt{2}\log k
\end{split}
\ee

Finally, in these settings the generalised cusp anomalous dimension reads

\be
\Gg_{\rm cusp}^{AdS}(\gl,\phi\to\infty,\vartheta\to 0) = -\ii\sqrt{\gl}\left[ k \mathbb{K}(k^2) + \frac{1}{k} \mathbb{E}(k^2)  \right]= 2\sqrt{\gl}\log k = \frac{\phi}{2}\,\sqrt{2\gl}
\ee

as was expected by its four-dimensional counterpart \cite{Kruczenski:2002fb}.


\section{Conclusions}

We have considered a pure gauge cusped Wilson loop in Minkowski  spacetime slightly taken off of the light-cone for $\cN=6$ SCS theory in light-cone gauge. In perturbation theory, we have shown that its expectation value reproduces the known results for the cusp anomalous dimension of ABJM at the second order in the 't~Hooft coupling $\gl=\frac{N}{k}$

\be
\Gg_{\rm cusp} = \gl^2
\ee

being $N$ the rank of the gauge groups and $k$ the Chern-Simons level. At the non-perturbative level, we solved the Bethe-Salpeter equation with a one-loop corrected kernel and found, for large values of the coupling $\gl$, an $\exp\sqrt{\gl\phi}$ exponential behaviour which has the wrong dependence on the cusp angle $\phi$ to be in agreement with the predicted $\exp\sqrt{\gl}\phi$ of AdS/CFT. This fact can be due to a variety of issues. First, the Bethe-Salpeter kernel ignores an infinite number of gauge interactions that arise at higher order of perturbation theory through effective couplings of the type depicted in Figure \ref{fig:highervertex}.  Secondly, although the perturbative computation suggests it is, at strong coupling the $a\to 0$ or $\phi\to\infty$ limit might not be  a smooth map between the superconnection  (\ref{superconnection}) and the purely bosonic operator. In this case other contributions might be relevant. \\

To further investigate this fact it would be interesting to analyse the behaviour of the stringy solution when the internal angle $\vartheta\to\pi$. This should suppress scalar and fermionic contributions and single out the gauge one. \\

A complete computation starting from the Euclidean cusp is of course in order. Though it implies solving a Bethe-Salpeter equation for a quite cumbersome kernel that includes interaction three-vertices. Hopefully there might be at least a kinematical region in which this task can be accomplished.\\

Moreover, the BS equation for corrected ladders could be analysed at weak coupling. It might be interesting to compute $\gl^4$ corrections to the cusp anomalous dimension and see whether they can tell something about the still evasive interpolating  function $h(\gl)$ \cite{Beisert:2005tm,Grignani:2008is,Nishioka:2008gz,Berenstein:2008dc,Berenstein:2009qd,
Minahan:2008hf,Gaiotto:2008cg}  through the conjecture made in \cite{Correa:2012at}.

\vspace{10mm}
{\bf Acknowledgements}

I would like to thank in particular Luca Griguolo for many advises and comments about this manuscript. I also acknowledge Domenico Seminara and Nadav Drukker for very useful discussions.


\newpage
\appendix

\section{Two-loops diagrams}
\label{sec:two-loops}

\begin{figure}[htb]
\begin{center}
\includegraphics[width=.12\textwidth]{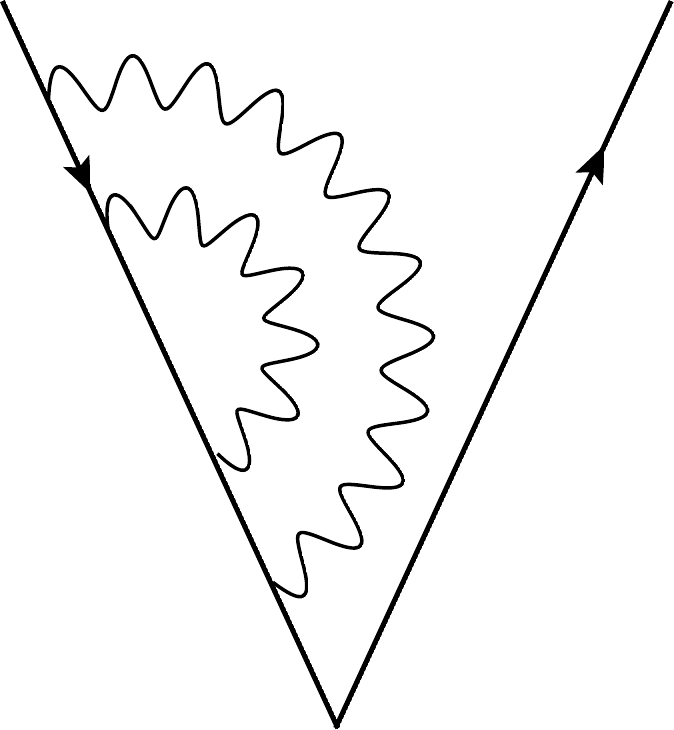}\mbox{(a)}
\includegraphics[width=.12\textwidth]{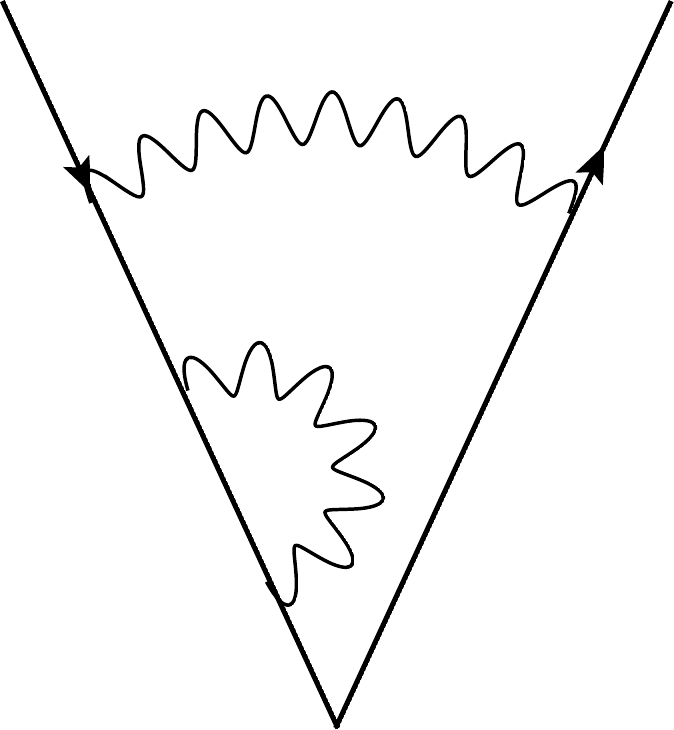}\mbox{(b)}
\includegraphics[width=.12\textwidth]{2gluons.pdf}\mbox{(c)}
\includegraphics[width=.12\textwidth]{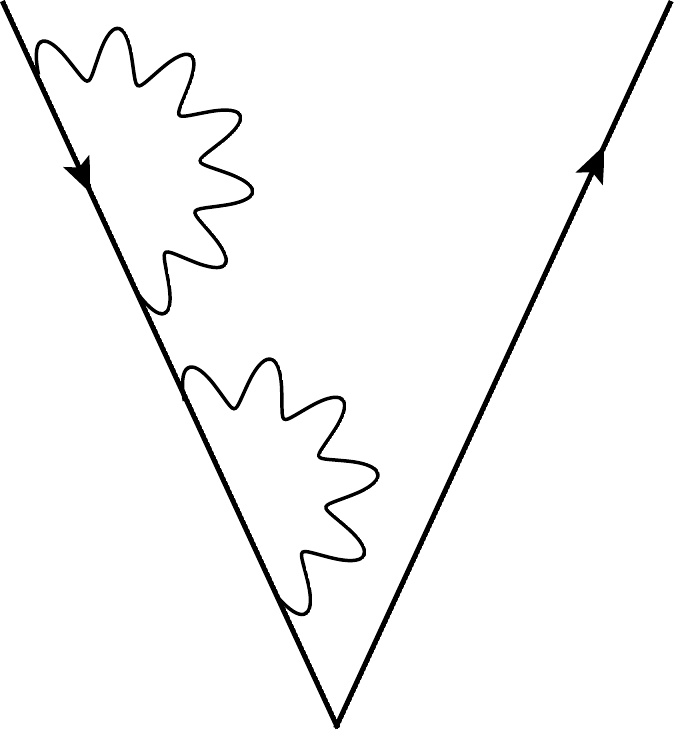}\mbox{(d)}
\includegraphics[width=.12\textwidth]{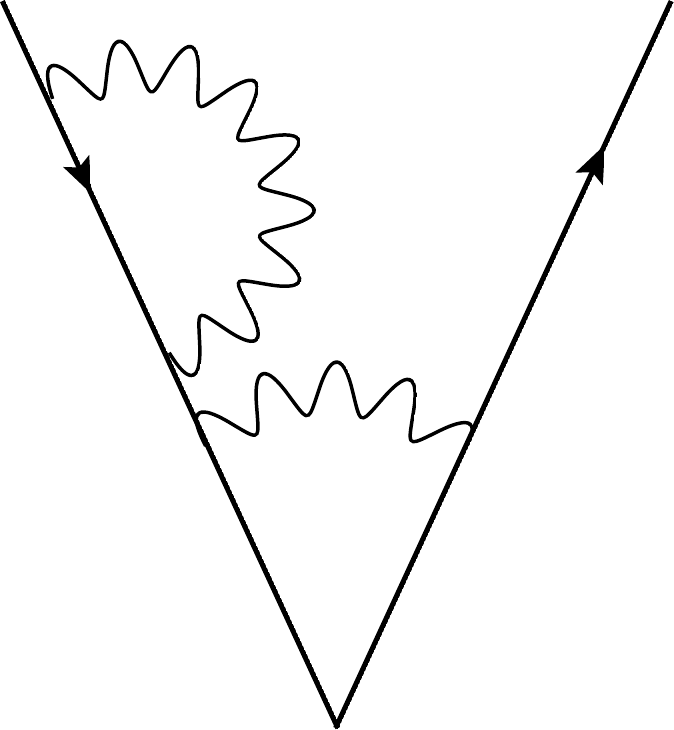}\mbox{(e)}
\includegraphics[width=.12\textwidth]{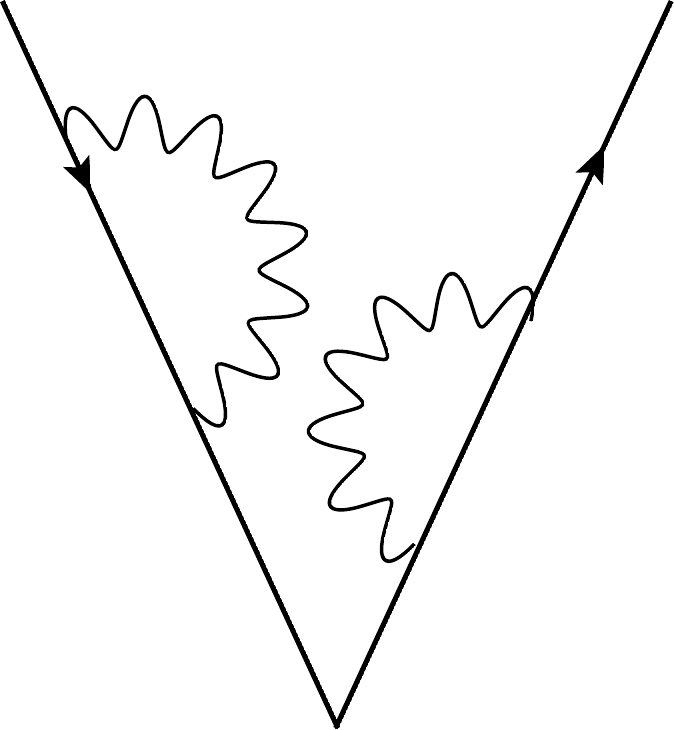}\mbox{(f)}
\includegraphics[width=.12\textwidth]{crossed_ladders.pdf}\mbox{(g)}
\includegraphics[width=.12\textwidth]{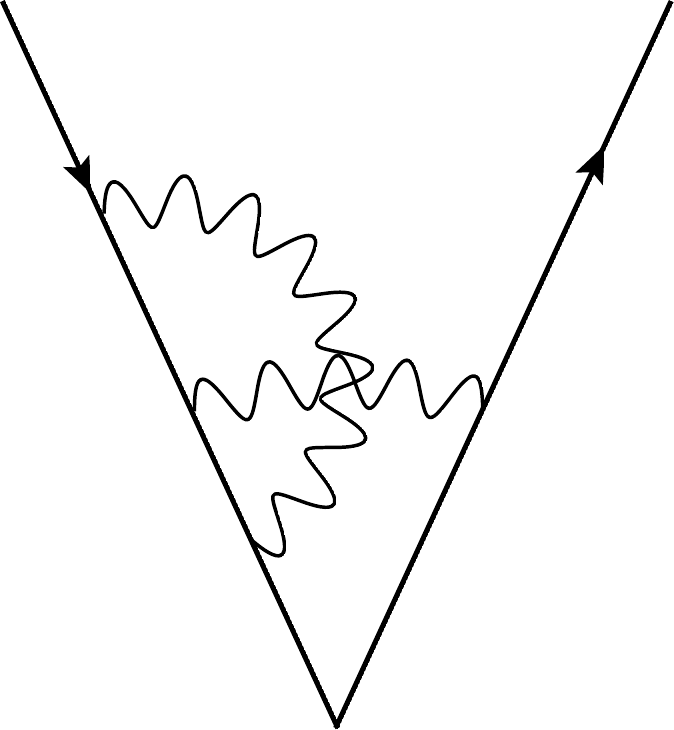}\mbox{(h)}
\includegraphics[width=.12\textwidth]{corrected_propagator.pdf}\mbox{(i)}
\includegraphics[width=.12\textwidth]{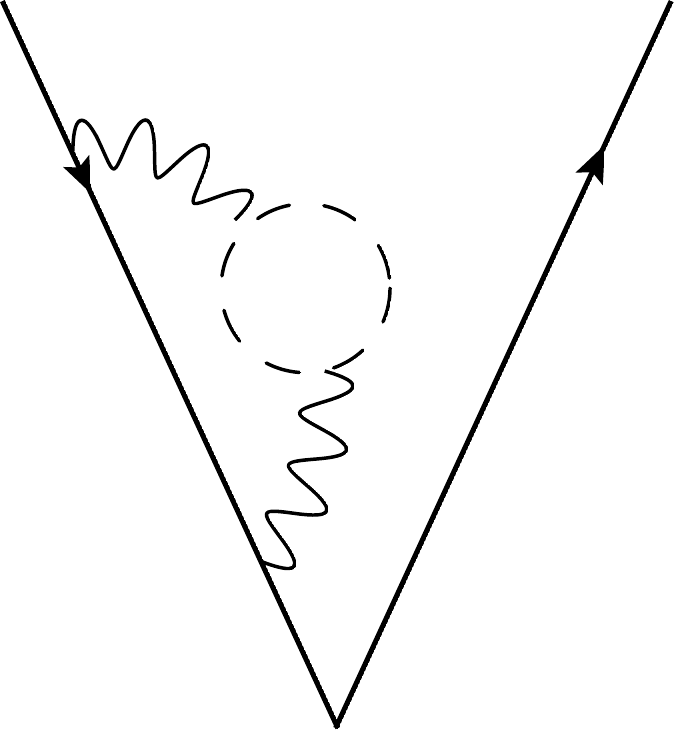}\mbox{(j)}
\includegraphics[width=.12\textwidth]{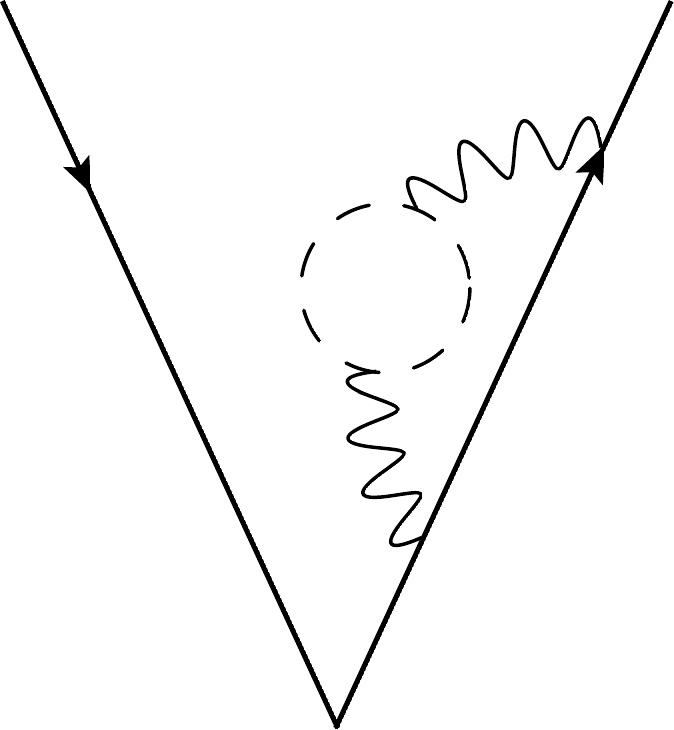}\mbox{(k)}
\includegraphics[width=.12\textwidth]{gluon-vertex.pdf}\mbox{(l)}
\end{center}
\caption[Graphs contributing at two-loops.]{\label{fig:all-two-loops} All the Feynman diagrams at the second order of the perturbative expansion of the Wilson loop in the Chern-Simons coupling $1/k$.}
\end{figure}

\subsection{On the light-cone}
\label{sec:on-lightlike}

Consider the perturbative expansion of the Wilson loop operator (\ref{eq:loopexpansion}) coupled to the strictly light-like contour $\cC$. At order $1/k^2$, all the possible contractions of the fields produce several Feynman diagrams,  depicted in Figure \ref{fig:all-two-loops}, though most of them are trivial. Indeed, any graph with at least one tree-level propagator whose both ends are stuck to same side of the cusp vanish for the antisymmetry of the propagator itself; so (a), (b), (d), (e), (f) are trivially zero. Also all non-planar diagrams, made exception for (g), vanish for the same reason. The latter, which would not contribute in any case to the large $N$ expansion, vanishes because the delta functions in the tree-level propagators can never be satisfied at the same time. Moreover the vertex diagram (l) does not even exist in light-cone gauge. At the end of the day the perturbative expansion reduces to the evaluation of the double exchange (c) and the corrected propagator (i), (j) and (k).\\

\vspace{10mm}
{\bf Double gauge field exchange}
~\\

Contracting the fourth order expansion of (\ref{eq:loopexpansion}) with the CS propagator produces double exchange graphs

\be
\label{eq:doubleexchangegraphs-2}
\int \dd s_i \,\dd s_j\, \dd s_k\, \dd s_l\, \dot x_i^m \dot x_j^n \dot x_k^r \dot x_l^s \bigg(\left< A_m A_n \right> \left< A_r A_s \right> + \left< A_m A_s \right> \left< A_n A_s\right> + \left< A_m A_r \right> \left< A_n A_s\right> \bigg)
\ee

In the first summand above, at least one propagator is stuck to same edge of the cusp, and hence vanishes for the same reason of the one-loop case. The third summand is a non-planar contribution and will be discussed below. The only relevant term is the second summand of (\ref{eq:doubleexchangegraphs-2}).  
Double propagators are easily seen to exponentiate the one loop result, namely exploiting the symmetry of the integrands one has

\be
\begin{split}
L^{({\rm ladder})}_{A} =&\frac{N^3}{k^2} \int_{-L}^{-\gep}\dd s_1\, \int_\gep^L \dd s_4\, G^{(1)}(s_1,s_4)
\int_{-s_1}^{-\gep}\dd s_2\, \int_\gep^{s_4} \dd s_3\, G^{(1)}(s_2,s_3) \\
=& \frac{N^3}{k^2} \half \left( \int_{-L}^{-\gep}\dd s_1\, \int_\gep^L \dd s_4\, G^{(1)}(s_1,s_4) \right)^2\\
=& \frac{N^3}{ 2 k^2}  \left( \log\frac{L}{\gep} \right)^2
\end{split}
\ee

Non-planar graphs, as the one depicted in Figure \ref{fig:twoloopsgraphs} (b) aren't of any worry when one is considering the planar limit of the theory.  On the other hand it can be shown that crossed diagrams do not contribute at all, even for finite $N$ and $M$. As a matter of fact, since the delta functions in the propagators restrict the support of integration to a region of order $\gep^2$ as $\gep\to 0$, while the divergence of the integrand is only logarithmic, the contributions are at most finite. In more details the crossed-ladder diagram reads

\be
\label{Lcrossed}
\begin{split}
L^{({\rm crossed})} =& \frac{N}{k^2} \int_{-L}^{-\gep}\dd s_1\, \int_\gep^L \dd s_3\, G^{(1)}(s_1,s_3)
\int_{-s_1}^{-\gep}\dd s_2\, \int_{s_3}^L \dd s_4\, G^{(1)}(s_2,s_4) \\
=&  \frac{N}{k^2} \int_{-L}^{-\gep}\dd s_1\, \int_\gep^L \dd s_3\, \frac{\left(2\dot x_1^+ \dot x_3^+ \dot x_1 \cdot \dot x_3  \right)^\half}{\dot x_1^+ s_1 -\dot x_3^+ s_3}\gd(\dot x_1^T s_1 - \dot x_3^T s_3)\\
& \times \int_{s_1}^{-\gep}\dd s_2\, \int_{s_3}^L \dd s_4\,\frac{\left(2\dot x_2^+ \dot x_4^+ \dot x_2 \cdot \dot x_4  \right)^\half}{\dot x_2^+ s_2-\dot x_4^+ s_4}\gd(\dot x_2^T s_2 - \dot x_4^T s_4)\\
=&  \frac{N}{k^2} \int_\gep^L \dd s_3\, \frac{1}{s_3} \Theta\left( s_2 -\frac{s_3}{\rho}  \right) \Theta\left(L+ \frac{s_3}{\rho} \right)\\
&\times \int_{-L}^{-\gep} \dd s_2\, \frac{1}{s_2} \Theta\left( \rho s_2 -s_3 \right) \Theta\left( L-\rho s_2 \right)
\end{split}
\ee

where the $\Theta$'s emerge in a careful treatment as existence conditions for the solutions of delta functions and $\rho=\frac{\dot x_2^T}{\dot x_4^T} = \frac{\dot x_1^T}{\dot x_3^T}$. So the computation of the crossed ladder diagrams amounts to the computation of the simple integral $\int_{\Omega} 1/(s_2 s_3)$ on the subset $\Omega$, that in turn is determined by the intersection of the conditions

\be
\begin{split}
& \rho>0 \quad \Rightarrow \quad \{\rho s_2 >s_3\} \cup \{-L<s_2<-\gep\} \cup \{\rho s_2<L\} \cup \{s_3>-\rho L\} =\emptyset \\
& \rho<0 \quad \Rightarrow \quad \{\rho s_2 >s_3\} \cup \{-L<s_2<-\gep\} \cup  \{s_3<-\rho\} \cup \{\rho s_2< s_3\} =\emptyset
\end{split}
\ee

which is the empty set. One might argue that introducing a framing vector would give rise to a non empty intersection of the conditions above. It is straight forward to add a framing vector to (\ref{Lcrossed}) above

\be
\label{Lcrossedframed}
\begin{split}
L^{({\rm crossed})}_\eta =& \int_{-L}^{0}\dd s_1\, \int_0^L \dd s_3\, \frac{\left(2\dot x_1^+ \dot x_3^+ \dot x_1 \cdot \dot x_3  \right)^\half}{\dot x_1^+ s_1 -\dot x_3^+ s_3 +\eta_{13}}\gd(\dot x_1^T s_1 - \dot x_3^T s_3 + \eta_{13})\\
& \times \int_{s_1}^{0}\dd s_2\, \int_{s_3}^L \dd s_4\,\frac{\left(2\dot x_2^+ \dot x_4^+ \dot x_2 \cdot \dot x_4  \right)^\half}{\dot x_2^+ s_2-\dot x_4^+ s_4+\eta_{24}}\gd(\dot x_2^T s_2 - \dot x_4^T s_4+\eta_{24})\\
=& \int_0^L \dd s_3\, \frac{1}{s_3+\eta_{13}} \Theta\left( s_2 -\frac{s_3-\eta_{13}}{\rho}  \right) \Theta\left(L+ \frac{s_3-\eta_{13}}{\rho} \right)\\
&\times \int_{-L}^{0} \dd s_2\, \frac{1}{s_2+\eta_{24}} \Theta\left( \rho (s_2 +\eta_{24})-s_3 \right) \Theta\left( L-\rho (s_2+\eta_{24}) \right)
\end{split}
\ee

with a slight abuse of notation. Note that the framing procedure regularises UV divergences due to propagators contracting at the cusp point, making the regulator $\gep$ irrelevant. Now theta function constraints do admit solutions, but it is easy to see that the domain of integration squeezes at least as $\cO(\eta)$ as $\eta\to 0$, whereas the integrand is at most logarithmically divergent. So we can safely assume that $L^{({\rm crossed})}_\eta$ is at most finite. \footnote{In that case it would be tantalising to assume it is a framing contribution, but this belief would be in sharp contrast with the notion of framing contributions as coming from graphs with contractible propagators.}

\vspace{10mm}
{\bf One-loop corrected gauge field propagator}
~\\

The propagator of the gauge field corrected at one-loop order, Figure \ref{fig:all-two-loops} (i), emerges from the contraction of the quadratic piece of the expansion  (\ref{eq:loopexpansion}) with two powers of the interaction Lagrangian 

\be
 - \int \dd s_i\,\dd s_j \, \dot x_i^m \dot x_j^n \, A_m A_n\,\half \left( \ii\int \dd^3 y \cL_{\rm int}(y) \right)^2 = \half \int \dd s_i\,\dd s_j \, \dot x_i^m \dot x_j^n G_{mn}^{(2)}(x_i,x_j)
\ee 

Using the results of Section (\ref{sec:corrprop}) one can show, with some manipulation, that the integrand above can be written as the sum of the one-loop corrected propagator computed in Landau gauge in \cite{Drukker:2008zx} plus a semi-total derivative term

\be
\label{eq:twoloopint1}
\dot x_1^m \dot x_2^n G_{mn}^{(2)}(x_1,x_2)=
 \frac{N^3}{k^2}\left[ -\frac{\dot x_1 \cdot \dot x_2 }{(x_1-x_2)^2} + \half\left( \dot x_2^+ \frac{d}{ds_1} - \dot x_1^+ \frac{d}{ds_2} \right) \frac{1}{[x^+]}\log\left( -\frac{x^2}{(x^T)^2} \right) \right]
\ee

It is convenient to split the domain of integration and consider separately the three graphs in Fig.\ref{fig:all-two-loops} (i), (j), (k). The first of them is the sum of the following three integrals

\be
L^{(2)}_{\rm vector} = -\int_{-L}^{-\gep} \dd s_1\, \int_\gep^L \dd s_2\,  \frac{\dot x_1 \cdot \dot x_2 }{(x_1-x_2)^2} = -\int_{\gep}^{L} \dd s_1\, \int_\gep^L \dd s_2\,\frac{\dot x_1 \cdot \dot x_2 }{2 \dot x_1 \cdot \dot x_2 s_1 s_2}= - \half\left( \log\frac{L}{\gep} \right)^2
\ee

\be
\begin{split}
L^{(2)}_{\rm ext1} =& \half \int_{-L}^{-\gep} \dd s_1\, \int_\gep^L \dd s_2\, \dot x_2^+ \frac{d}{ds_1} \frac{1}{[x^+]}\log\left( -\frac{x^2}{(x^T)^2} \right)
\end{split}
\ee

\be
\begin{split}
L^{(2)}_{\rm ext2} =& -\half \int_{-L}^{-\gep} \dd s_1\, \int_\gep^L \dd s_2\, \dot x_1^+ \frac{d}{ds_2} \frac{1}{[x^+]}\log\left( -\frac{x^2}{(x^T)^2} \right)
\end{split}
\ee

They are all standard integrals which are completely prescribed by the UV and IR cutoffs $\gep$ and $L$. The first one was trivial, the second reads

\be
\begin{split}
L^{(2)}_{\rm ext1}=& \half \left[ -\log(s_2)\,\log\left(1-\frac{\dot x_2^+ s_2}{\dot x_1^+ s_1}\right) -\log(\dot x_1^+ s_1-\dot x_2^+ s2)\,\log\left( \frac{2\dot x_1\cdot \dot x_2 s_1 }{(\dot x_2^T)^2}  \right) \right.\\
&+ 2\log\left(s-\frac{\dot x_1^T s_1}{\dot x_2^T}  \right)\, \log\left(-\frac{(\dot x_2^+ s_2- \dot x_1^+ s_1)\dot x_2^T }{(\dot x_1^T \dot x_2^+ - \dot x_1^+ \dot x_2^T)s_1} \right) - Li_2\left( \frac{\dot x_2^+ s2}{\dot x_1^+ s_1} \right) \\
&\left. + 2Li_2 \left( \frac{(\dot x_2^+ s_2- \dot x_1^+ s_1)\dot x_2^T }{(\dot x_1^T \dot x_2^+ - \dot x_1^+ \dot x_2^T)s_1} \right) \right]_{s_1=-L,\,s_2=\gep}^{s_1=-\gep,\,s_2=L} \\
=& \frac{1}{4} \left( \log\frac{L}{\gep} \right)^2 - \half \log\frac{L}{\gep} \log\frac{(-2 \dot x_1 \cdot \dot x_2)}{(\dot x_1^T)^2} + \cO(1)
\end{split}
\ee

From the symmetries of the problem it is clear that $L^{(2)}_{\rm ext1}$ and $L^{(2)}_{\rm ext2}$ give the same result up to interchanging $\dot x_1 \leftrightarrow \dot x_2$, hence diagram \ref{fig:all-two-loops} (i) accounts for

\be
I^{(2)}_{(i)} = -\half\frac{N^3}{k^2}  \log\frac{L}{\gep} \log\left(\frac{-2 \dot x_1 \cdot \dot x_2}{\dot x_1^T \dot x_2^T} \right)
\ee

Differently from the four-dimensional case, single-leg graphs of Fig. \ref{fig:all-two-loops} (j) and (k) do not vanish, instead they are crucial in reconstructing the correct result. Treating (\ref{eq:derivatives}) carefully, one can see that the first term drops and the relevant integrands in the present case read

\be
\half\left( \frac{d}{ds_1} - \frac{d}{ds_2} \right) \frac{\dot x_i^+}{[\dot x_i^+(s_1-s_2)]}\log\left( -\frac{[\dot x_i(s_1-s_2)]^2}{[\dot x_i^T(s_1-s_2)]^2} \right)
\ee

where the Mandelstam-Leibbrandt prescription on the spurious poles of the propagator removes any potential contact divergence. The latter is indeed equivalent in this case to a framing prescription  by means of a small vector orthogonal to the contour  $|\eta|\sim 0$ and $\eta\cdot x_i(s)=0$. Then the $\gep$ cutoff can be removed, as contact divergences, including the one in $s=0$ that arises when the two endpoints of the propagator pinch each other in the origin, are correctly regularized by the ML prescription on the propagator. Hence the two integrals in Fig. \ref{fig:all-two-loops} (j) and (k) can be conveniently merged into a single integral; to this end note that, since each of them only depends on the square of one of the $\dot x_i$'s, each integrand is separately even if both sign of $s$'s are reversed \footnote{Reversing the sign of the parameters $s_{1,2}$ actually accounts for a flip in the sign of the framing vector $\eta$. Though this does not affect the final result, being divergences logarithmic in the moduls of $\eta$.}

\be
\begin{split}
\label{eq:singleleg-2}
I^{(2)}_{(j)} =&  \int_{-L}^{0} \dd s_1\, \int_{s_1}^{0} \dd s_2\, \half\left( \frac{d}{ds_1} - \frac{d}{ds_2} \right) \frac{1}{s_1-s_2+\eta}\log\left( -\frac{x_1^2}{(x_1^T)^2} \right)\\
=& \int_{0}^{L} \dd s_1\, \int_{0}^{s_1} \dd s_2\, \half\left( \frac{d}{ds_1} - \frac{d}{ds_2} \right) \frac{1}{s_1-s_2-\eta}\log\left( -\frac{x_1^2}{(x_1^T)^2} \right)\\
\end{split}
\ee

so that it can be merged with $I^{(2)}_{(k)}$

\be
\begin{split}
I^{(2)}_{(j)}+I^{(2)}_{(k)} 
=& \int_{0}^{L} \dd s_1\, \int_{0}^{L} \dd s_2\, \half\left( \frac{d}{ds_1} - \frac{d}{ds_2} \right) \frac{1}{s_1-s_2+\eta}\log\left( -\frac{x_1^2}{(x_1^T)^2} \right)
\end{split}
\ee

The contribution of the two derivatives is again the same, let us consider the first global integral

\be
\begin{split}
& \half \int_{0}^{L} \dd s_1\, \int_{0}^{L} \dd s_2\, \frac{d}{ds_1} \frac{1}{s_1-s_2+\eta}\log\left( -\frac{x_1^2}{(x_1^T)^2} \right)\\
=& -\left(\log\frac{L}{\eta} \right)^2 - \log\frac{L}{\eta}\,\log (x_1^T)^2 + \cO\left( \frac{\eta}{L} \right)
\end{split}
\ee

Had we used the framing prescription instead of the cutoff $\gep$ in all integrals, we would have recovered the same results, up to the simple identification of $\gep\leftrightarrow\eta$. It clearly follows from the fact that the framing simply acts as a cutoff for the contact divergence located at $s=0$. We will use this identification throughout the computations.


\subsection{Near the light-cone}
\label{sec:on-deformed}

We want now to analyse the perturbative expansion of the Wilson loop coupled to the deformed contour $\cC_{\rm def}$. Diagrams contributing to this case are the same of the strictly light-like case. We will not consider the counterpart of the double exchange in Fig.\ref{fig:all-two-loops} (a) as the one loop analysis of (\ref{eq:LAreg}) clarifies that the topological nature of the CS field is such that the dependence on the velocity (cusp angle) $\phi$ is suppressed. We will therefore use the results obtained in the light-like case.  Let us precede with the computation of the contribution coming from the corrected CS propagator. Diagram (i) of Fig.\ref{fig:all-two-loops} reads in this case (using $\ga = \sin\gth$)

\begin{multline}
\int_{-L}^{-\gep} \dd s_1\, \int_\gep^L \dd s_2\, \left[ -\frac{\dot x_1 \cdot \dot x_2}{ 2a\ga (s_1^2-s_2^2) -2(\dot x_1 \cdot \dot x_2)s_1 s_2} \right. \\
\left. + \half\left( \dot x_2^+ \frac{d}{ds_1}- \dot x_1^+ \frac{d}{ds_2}\right) \frac{1}{\dot x_1^+ s_1-\dot x_2^+ s_2}\log\left( -\frac{2a\ga (s_1^2-s_2^2) -2(\dot x_1 \cdot \dot x_2)s_1 s_2}{[\ga(s_1-s_2)-a(s_1+s_2)]^2} \right) \right]
\end{multline}

and can be decomposed as before in the sum of a vector and a semi-total derivative parts. After changing variable to $s_1\to -Ls,\, s_2\to Lxs$, the former integral reads

\be
\begin{split}
& -\int_\gep^1 \dd s\, \frac{1}{s}\int_{\gep/s}^{1/s} \dd x\, \frac{\dot x_1 \cdot \dot x_2}{ 2a\ga (1-x^2) -2(\dot x_1 \cdot \dot x_2)x}\\
=& -\int_\gep^1 \dd s\, \frac{1}{s}\int_{\gep}^{1} \dd x\, \frac{\dot x_1 \cdot \dot x_2}{ 2a\ga (1-x^2) -2(\dot x_1 \cdot \dot x_2)x}\\
&- \int_{\gep}^{1} \dd x\, \frac{\dot x_1 \cdot \dot x_2}{ 2a\ga (1-x^2) -2(\dot x_1 \cdot \dot x_2)x} \int_{1/x}^1 \dd s\, \frac{1}{s}\\
=& -\int_\gep^1 \dd s\, \frac{1}{s} \frac{y}{\sqrt{-1-y^2}}\left[  {\rm arctan}\left( \frac{y+1}{\sqrt{-1-y^2}}\right)-{\rm arctan}\left( \frac{y+\gep}{\sqrt{-1-y^2}} \right)\right]\\
&+\frac{1}{2\sqrt{y^2+1}} \left[ \text{Li}_2 \left(-\frac{x}{y-\sqrt{y^2+1}}\right)-\text{Li}_2 \left(-\frac{x}{y+\sqrt{y^2+1}}\right) \right. \\
& \left. +\log(x) \left(\log\left(\frac{x}{y-\sqrt{y^2+1}}+1\right)
- \log\left(\frac{x}{y+\sqrt{y^2+1}}+1\right)\right) \right]_{x=\gep}^{x=1}
\end{split}
\ee

where $y= \frac{\dot x_1 \cdot \dot x_2}{2a\ga}$ is taken to be very large and with a slight abuse of notation $\gep=\gep/L$. Then the second and third lines after the last equal sign above vanish and the only contribution that survives is the first integral expanded for $y\to\infty$

\be
= -\log{\frac{L}{\gep}} + \cO\left( \frac{\gep}{L} \right)
\ee

Each of the two semi-total derivative pieces can be written as

\be
\begin{split}
&\half \int_{-L}^{-\gep} \dd s_1\, \int_\gep^L \dd s_2\, \frac{d}{ds_1}\frac{1+\gb}{-s_1-s_2+\gb(s_1-s_2)} \log\left( \frac{2(2\gb^2-a^2)s_1 s_2 -2a\ga(s_1^2-s_2^2)}{[\ga(s_1-s_2)-a(s_1+s_2)]^2}  \right)\\
=& -\half \log\frac{L}{\gep}\,\log(a) + \cO\left( \frac{\gep}{L} \right)
\end{split}
\ee

Let us consider single-leg diagrams of Figure \ref{fig:all-two-loops} (j) and (k). As in the previous case, integrals must be consistently regularised using a framing  prescription, and once done that, the cutoff $\gep$ must be removed (it can be kept, but it would simply be redundant). Hence again integrals referring to different legs can be merged and we are left with the three global integrals

\begin{multline}
\int_0^L \dd s_1\, \int_0^L \dd s_2\,\left[- \frac{2a\ga}{2a\ga(s_1-s_2)^2 + \eta^2} \right. \\
\left. + \half \left( \frac{d}{ds_1}-\frac{d}{ds_2} \right) \frac{1}{s_1-s_2+\eta} \log\left( \frac{2a\ga(s_1-s_2)^2+\eta^2}{[(\ga-a)(s_1-s_2)]^2}  \right) \right]
\end{multline}

The first integral is straightforward

\be
-\int_0^L \dd s_1\, \int_0^L \dd s_2\,\frac{2a\ga}{2a\ga(s_1-s_2)^2 + \eta^2} 
= 2\log\frac{L}{\eta} + \log(a) + \cO\left( \frac{\gep}{L} \right)
\ee

and the two pieces that are left again produce the same contribution

\be
\begin{split}
& \half \int_0^L \dd s_1\, \int_0^L \dd s_2\, \frac{d}{ds_1}\frac{1}{s_1-s_2+\eta} \log\left( \frac{2a\ga(s_1-s_2)^2+\eta^2}{[(\ga-a)(s_1-s_2)]^2}  \right)\\
=& \log\frac{L}{\eta}\, \log(a) + \cO\left( \frac{\gep}{L} \right)
\end{split}
\ee


\section{Chern--Simons theory in lightcone gauge}
\label{sec:csaction}

Given a semisimple compact Lie group $G$ and a connection $A$, the well known Chern--Simons action in 3--dimensional Minkowski spacetime reads

\be
S_{CS} = \frac{k}{4 \pi} \int_{M^3} \Tr\left[A\wedge dA + \frac{2}{3}A\wedge A \wedge A\right] d^3x
\ee

being $k$ the "Chern-Simons" level and $\Tr$ the trace over the fundamental representation of $G$. For any continuous map $h:M^3\to G$ which is connected to the identity, the CS action is invariant under the gauge transformation

\be
\label{eq:csgaugetransf}
A^\mu \to h^{-1}A^\mu h+ h^{-1}\d^{\mu}h
\ee 

If $h$ is not connected to the identity the action gets added a term proportional to the winding numer $w$ of the map $h$, namely $S_{CS}\to S_{CS}^{(h)}=S_{CS}+2\pi k w$, so if $k$ is also an integer, the path-integral is left untouched and (\ref{eq:csgaugetransf}) is again a gauge transformation. Henceforth $k$ is said to be quantized to an integer.

For a $SU(N)$ gauge connection $A_\mu=A_\mu^a T^a$ we use the conventions 

\be
\left[T^a,T^b \right]=f_{abc}T^c, \qquad \Tr(T^a T^b) = \half \gd^{a,b}
\ee

for the generators in the fundamental representation.
Let us introduce a null vector $n^2=0$, the full action including the homogeneous light-cone gauge fixing $n^\mu A_\mu=0$ reads in components

\be
S_{CS} = \int\dd^3 x\, \gep^{\mu\nu\rho} \left[ \half A_\mu^a \d_\nu A_\rho^a + \frac{g}{3!} f^{abc} A_\mu^a A_\nu^b A_\rho^c \right]-\frac{1}{2\ga}(n^\mu A_\mu^a)^2 +\bar c^a n^\mu D_\mu^{ab}c^b
\ee 

where we have rescaled $A \to g A = \sqrt{4\pi/k}A$ to make sense of the perturbative expansion. The fields $c,\bar c$ are ghost fields and the covariant derivative acts as $D_\mu^{ab} =\d_\mu \gd^{ab} +g f^{abc}A_\mu^c$. It is understood that the light-cone gauge condition is imposed in the limit $\ga\to 0$.

\n It is well known in the literature that lightcone gauge computations are usually easier than in other gauges, indeed much of the seminal work about knot invariants and Chern-Simons theory widely exploited this gauge. The main draw back is the obvious loss of Lorentz covariance in computations, though physical observables preserve covariance.

It is convenient to shift to light-cone coordinates from now on, that we indicate with
Roman lower-case letters $m,n,r ...$ to not confuse them with Cartesian indices $\mu,\nu,\rho ...$. The new indices take value $m=+,-,T$, which correspond to 

\be
A_+=A_0+A_1, \qquad A_-=A_0-A_1, \qquad A_T=A_2
\ee 

Picking up the light-cone vector $n^\mu=(1,1,0)$, the gauge fixing 

\be
A_m n^m = 0
\ee

imposes

\be
A_-=0
\ee

The power of this gauge choice reseeds in the fact that now the ghosts decouple from the gauge fields and the interaction vertex, being totally antisymmetric in the indices $+,-,T$, boils down to zero. The bottom line is that the actions takes the rather simple form of a free-field action

\be
\begin{split}
\label{eq:CSactionlightcone}
& S_{CS} = \int \dd^3x\,\gep^{mn}A_m^a \d_{-} A_n^a\\
& S_{gh} = \int \dd^3x\,\bar c^a \d_{-} c^a 
\end{split}
\ee

where the indices now run over $+,T$ only and $\gep^{+T}=-\gep^{T+}=1$. It is well known that the CS action in covariant gauges receives a one-loop contribution which shifts the level $k$ of an amount equal to the quadratic Casimir of the adjoint representation of the gauge group $k\to k+ {\rm sign}(k)C_A^2$ \cite{Guadagnini:1989kr} \cite{AlvarezGaume:1989wk} \cite{Giavarini:1992xz}. Moreover no two-loops corrections are present and higher loops are ruled out by power-counting, consistently with the non-perturbative analysis of \cite{Witten:1988hf}. It was proved in \cite{Leibbrandt:1991qr} \cite{Leibbrandt:1993mr} that the same shift at one-loop also occurs in axial type gauges, once the unphysical poles introduced by the gauge choice are consistently regularised, and this is believed to hold at higher loops as for the case of covariant gauges. 

For later convenience note that some attention to indices is required, namely we use the conventions in which the metric reads

\be
\label{eq:lc-metric}
\eta^{mn}=\eta_{mn}=(U^{-1})_{m\mu}\, g^{\mu\nu}\, U_{\nu n}=\left(
\begin{array}{ccc}
 0 & 1 & 0 \\
 1 & 0 & 0 \\
 0 & 0 & -1\\
\end{array}\right)
\ee

where $g={\rm diag}(1,-1,-1)$ is the Minkowski metric. This is consistent with the choice

\be
\begin{split}
& x_+=\frac{1}{\sqrt 2}\left(x_0+x_1\right), \qquad x_-=\frac{1}{\sqrt 2}\left(x_0-x_1\right), \qquad x_T=x_2\\
& k_+=\frac{1}{\sqrt 2}\left(k_0+k_1\right), \qquad k_-=\frac{1}{\sqrt 2}\left(k_0-k_1\right), \qquad k_T=k_2\\
\end{split}
\ee

in particular note that 

\be
x_+=x^-, \qquad x_-=x^+, \qquad x_T=-x^T, \qquad x^2=2x_+x_- - x_T^2
\ee

and 

\begin{multline}
k\cdot x= k_m \eta^{mn} x_n = k_+ x_- + k_- x_+ - k_T x_T =\\
= k^+x_+ +k^-x_- +k^Tx_T = k_0 x_0- k_1 x_1 - k_2 x_2
\end{multline}

so that the rule of summing contracted indices is preserved.


\subsection{Tree-level propagator and the Mandelstam-Leibbrandt prescription}

According to (\ref{eq:CSactionlightcone}), to compute the propagator one has to invert the operator $\gep^{mn}\d_-$, which conveniently written in momentum space reads $-\ii \gep^{mn} p^+$; this yields to

\be
\label{eq:treelevelgaugeprop}
\left< A_m^a A_n^b \right> =  \frac{1}{2}\gep_{mn}\gd^{ab}\frac{1}{p^+}
\ee

The spurious pole $p^+=0$ is to be shifted according to the Mandelstam-Leibbrandt causal prescription \cite{Mandelstam:1982cb} \cite{Leibbrandt:1983pj}

\be
\label{eq:ML}
[p^+]=p^+ + \ii\gep \, {\rm sign}(p^-)
\ee

which guarantees that the propagator is a genuine tempered distribution \cite{Bassetto:1992gv} (opposed to the Cauchy principal value prescription). The consistency of such prescription on the spurious poles caused by the non-covariant gauge choice was also pointed out in \cite{Bassetto:1984dq} in the case of Yang-Mills theory in four dimensions. Here $\gep$ is a small positive quantity. Let us recall that the antisymmetric tensor is now two-dimensional, hence the factor of a half. To reach a coordinate space expression we must evaluate the contour integral

\be
\int_{-\infty}^{+\infty} \frac{\dd p^+ \dd p^- \dd p^T}{(2\pi)^3} \frac{\e^{-\ii p^+ x_+ -\ii p^- x_- -\ii p^T x_T} }{p^+ + \ii\gep \, {\rm sign}(p^-)}
\ee

Note that we explicitly avoid using dimensional regularization. The integral over $p^T$ is trivial

\be
\gd(x_T) \, \int_{-\infty}^{+\infty} \frac{\dd p^+ \dd p^- }{(2\pi)^2} \frac{\e^{-\ii p^+ x_+ -\ii p^- x_-} }{p^+ + \ii\gep \, {\rm sign}(p^-)}
\ee

The integral over $p^+$ is straightforwardly carried out summing the residues of the integrand. Indeed when $p^- >0$, the $p^+=0$  pole in the complex $p^+$ plane is shifted below the real axes and we can close the contour with a big semi-circle enclosing the lower half of the complex plane 

\be
\int_{p^->0} \frac{\dd p^+ \dd p^- }{(2\pi)^2} \frac{\e^{-\ii p^+ x_+ -\ii p^- x_-} }{p^+ + \ii\gep} = -\ii \int_0^\infty \frac{\dd p^- }{2\pi} \e^{-\ii \gep x_+ -\ii p^- x_-} =
-\ii \int_0^\infty \frac{\dd p^- }{2\pi} \e^{-\ii p^- (x_- -\ii\gep' x_+)}
\ee

where we have traded $\gep$ for $\gep' =  \gep /|p^-|$ and an extra minus sign arises from the clockwise orientation of the contour. In the case where $p^-<0$ the contour can be closed encircling (counter-clockwise) the upper half complex plane and the result is analogous

\be
\int_{p^-<0} \frac{\dd p^+ \dd p^- }{(2\pi)^2} \frac{\e^{-\ii p^+ x_+ -\ii p^- x_-} }{p^+ - \ii\gep} = \ii \int_{-\infty}^0 \frac{\dd p^- }{2\pi} \e^{-\ii p^- (x_- -\ii\gep' x_+)}
\ee

Note that there is a hidden minus sign in $\gep'$ above due to the fact that $p^-\gep'={\rm sign}(p^-)\gep$. Summing the two integrals one has

\be
\begin{split}
& \ii \int_{-\infty}^0 \frac{\dd p^- }{2\pi} \e^{-\ii p^- (x_- -\ii\gep' x_+)}-\ii \int_0^\infty \frac{\dd p^- }{2\pi} \e^{-\ii p^- (x_- -\ii\gep' x_+)}=\\
& =-\frac{1}{2\pi} \frac{\gth(x_+)}{x_- +\ii\gep x_+}+\frac{1}{2\pi} \frac{\gth(-x_+)}{x_- -\ii\gep x_+}=\\
& =-\frac{1}{2\pi} \frac{1}{x_- +\ii\gep\,{\rm sign}(x_+)} =-\frac{1}{2\pi} \frac{1}{[x_-]}
\end{split}
\ee

where it is understood that the first integral only converges for $x_+>0$, whereas the second for $x_+<0$. Quite interestingly the ML prescription on the spurious poles of the momentum space propagator correctly prescribes the spurious poles in the coordinate space. At the end of the day the CS field propagator reads

\be
\label{eq:CSprop-lightcone}
\left< A_m^a(x) A_n^b(y) \right> =G_{mn}^{ab}(x-y)= -\frac{1}{4\pi}\gep_{mn}\gd^{ab} \frac{\gd(x_T-y_T)}{[x_- -y_-]}
\ee

As was already expected at the level of the light-cone  Lagrangean (\ref{eq:CSactionlightcone}), nothing propagates along the transverse coordinate and the dynamics is restricted to a two-dimensional subspace described by the light-cone directions. This analogy with two-dimensional dynamics was emphasised long ago in \cite{Frohlich:1989gr}; to this end note that the tree--level gauge field propagator is a solution of the Green equation

\be
\label{eq:csgreeneq}
\gep^{mn}\d_- G_{nr}^{ab}(x-y) =-\ii\gd(x-y)\gd^m_r \gd^{ab}
\ee

Actually, this equation is readly solved. First, the action (\ref{eq:CSactionlightcone}) has no derivative in the transverse direction, the propagator is then a delta function along $x_T$. Then
consider that the two-dimensional Feynman propagator is the Green function of the operator

\be
\left(\frac{\d}{\d x^0}\right)^2 - \left(\frac{\d}{\d x^1}\right)^2 =
\left(\frac{\d}{\d x^0} + \frac{\d}{\d x^1}\right) \left(\frac{\d}{\d x^0} - \frac{\d}{\d x^1}\right) = 2 \d_- \d_+
\ee

or otherwise stated \footnote{Note that the normalization of light-cone derivative is done according to $\d_m x^n=\gd_m^n$. As an example consider $\d_+ x^+ = \frac{1}{\sqrt 2}\left( \frac{\d}{\d x^0} - \frac{\d}{\d x^1} \right)\frac{1}{\sqrt 2}\left( x^0 - x^1\right) =1 $.}

\be
\label{eq:feynman}
2 \d_-\d_+ \Gd_{F}(x-y)= -\ii\gd(x-y)
\ee

\n so that 

\be
G_{mn}^{ab}(x)=-\ii\gd^{ab} \gep_{mn}\d_+ \Gd_{F}(x_+,x_-)\gd(x_T)
\ee

\n is a solution of (\ref{eq:csgreeneq}). The solution to the Green equation for the d-dimensional Laplace operator is well known in dimensional regularization

\be
\label{eq:feynmandimensional}
\Gd_{F}^{d}(x)=\int \frac{\dd^d p}{(2 \pi)^d}\, \frac{e^{-\ii p\cdot x}}{p^2} = \ii \Gg\left( \frac{d}{2}-1 \right)\frac{|x|^{2-d}}{4 \pi^{d/2}}
\ee

\n In $d=2+2\gd$ it exhibits a "dimensional divergence" plus logarithmic divergences in both the IR and UV. As a matter of fact consider the expansion of the Euler $\Gg(\gd)$ for $\gd\to 0$ 

\be
\Gg(\gd) = \ov{\gd}-\gg_E + \ov{6}\left( 3\gg_E +\frac{\pi^2}{2}\right) \gd +\cO(\gd^2)
\ee

\n so that (\ref{eq:feynmandimensional}) becomes

\be
\label{eq:feynmann2dim}
\Gd_{F}^{2+2\gd}(x) = \ov{2\pi\gd} -\ov{4\pi}\left( \log x^2 +\log\pi -\gg_E \right)+\cO(\gd) 
\ee

\n where $\gg_E$ is the Euler-Mascheroni constant. Hence the light-cone coordinate propagator reads

\be
\label{eq:prop-dim}
G_{mn}^{ab}(x)=-\frac{1}{4\pi} \gep_{mn} \gd^{ab} \frac{1}{x_-}\gd(x_T)
\ee

which is the same thing as in (\ref{eq:CSprop-lightcone}), up to a plausible prescription of spurious poles, which would naturally fall onto the ML prescription again. On the other hand the first derivation does not need to handle potentially harmful dimensional divergences and naturally takes track of the origin of the causal prescription. The Chern-Simons gauge field propagator geared with the ML prescription is well defined in strictly three dimensions.


\subsection{One-loop corrected gauge field propagator}
\label{sec:corrprop}

Choosing light-cone gauge fixing for the CS fields does not significantly alter the Lagrangean for 
fermions and scalars, beside the obvious fact that the covariant derivative along the gauged component of $A$ will be a standard derivative. At two-loops order the contribution to the gluon self energy due to ghost and gluon bubbles is known to cancel in covariant gauges. In light-cone gauge, ghosts decouple from the action and the CS triple vertex vanishes for antisymmetry, hence the gauge propagator receives contributions from the scalar and fermionic bubbles only. We denote these one-loop corrections as

\be
G^{(2)}_{mn}(p)= G^{(1)}_{mr}(p)\Pi^{rl}(p)G^{(1)}_{ln}(p)
\ee

\begin{figure}[htbf]
\begin{center}
\fbox{\includegraphics[width=.35\textwidth]{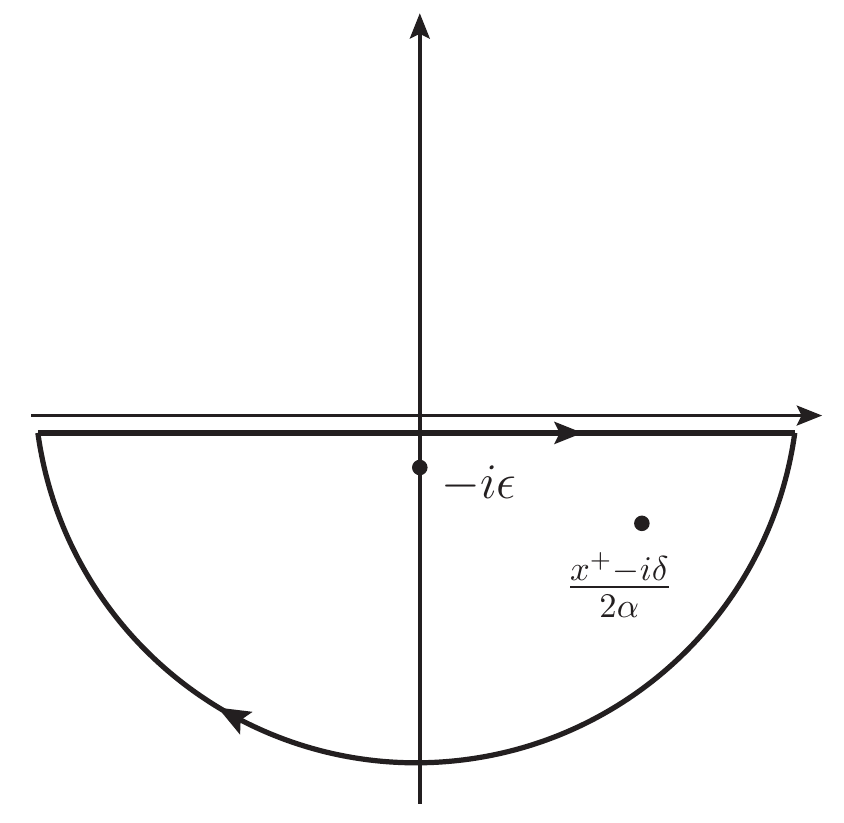}\mbox{(a)}}
\fbox{\includegraphics[width=.35\textwidth]{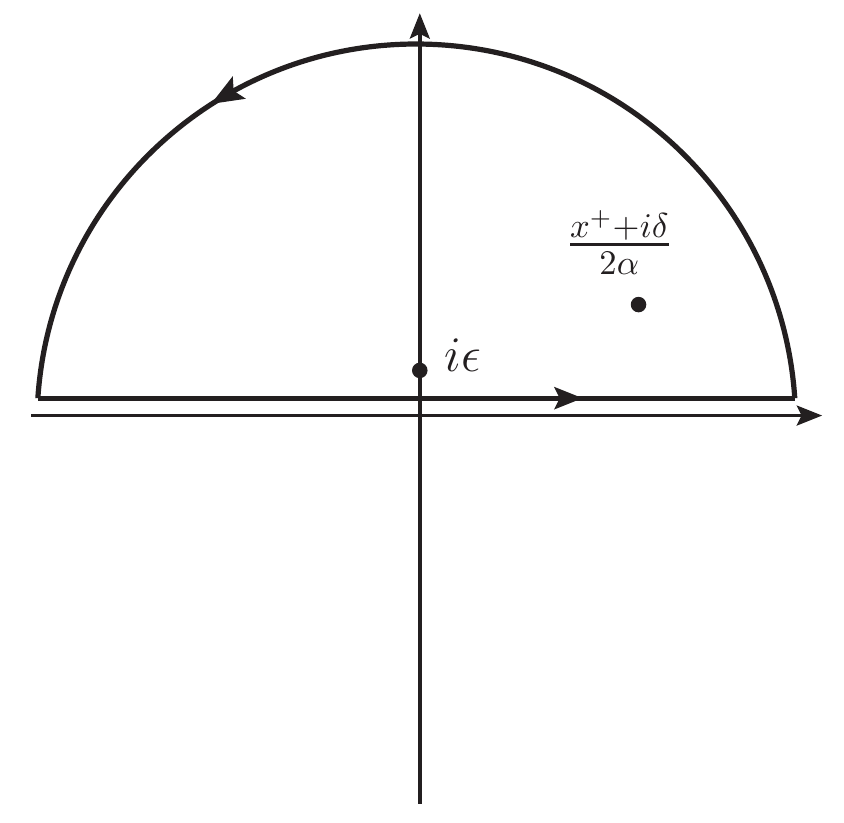}\mbox{(b)}}
\caption[Integration contours for the $p^+$ integral.]{\label{fig:p+intcontours} Integration contours for the $p^+$ integral. The position of poles related to the sign of $x^-$ is consistent with the Mandelstam-Leibbrandt prescription on the spurious poles in $x^+=0$. This shows how the ML prescription in momentum space is naturally reproduced in coordinate space.}
\end{center}
\end{figure}

Here $\Pi^{rl}$ is the sum of scalar and fermion loop contributions, it is just the same as in covariant gauges \cite{Drukker:2008zx}, and $ G^{(1)}_{mr}(p)$ was defined in (\ref{eq:treelevelgaugeprop}).  In $d=3-2\gep$ dimensions it reads

\be
\Pi_{\mu\nu}(p) = \gd^I_I \frac{2\ii N \mu^{2\gep}}{(4\pi)^{\frac{d}{2}}} \frac{\Gg\left(1-\frac{d}{2}\right) \Gg\left(\frac{d}{2}\right)^2}{\Gg\left(d-1\right)} \frac{p^2 g_{\mu\nu}-p_\mu p_\nu}{(-p^2)^{2-\frac{d}{2}}}
\ee

being $I=1,2,3,4$ the $R-$index of scalars and fermions. We only need to rotate its (cartesian) indices to light-cone ones

\be
\Pi^{rl} = (U^{-1})^{r \rho} \Pi_{\rho\gl} U^{\gl l}
\ee

where the matrix $U$ is implicitly defined in (\ref{eq:lc-metric}). Actually the action of $U$ on the indices structure above is trivial, so that the momentum space one-loop corrected gauge field propagator reads

\be
\label{eq:1loopgaugeprop}
G^{(2)\,ab}_{mn}(p) = -\left( \frac{2\pi}{k} \right)^2 \gd^I_I \frac{2\ii N \mu^{2\gep}}{(4\pi)^{\frac{d}{2}}} \frac{\Gg\left(1-\frac{d}{2}\right) \Gg\left(\frac{d}{2}\right)^2}{\Gg\left(d-1\right)} \frac{\gep_{mr}\gep_{ln} (p^2 \eta^{rl}-p^r p^l)}{[p^+]^2(-p^2)^{2-\frac{d}{2}}}
\ee

where all the latin indices take the only values $+$ and $T$. Having traded explicit Lorentz covariance for some spurious poles (a bargain), the Fourier integral

\be
\label{eq:uglyFT}
G^{(2)}_{mn}(x)= ({\rm factor}) \times \int_{-\infty}^{+\infty} \frac{\dd^3 p}{(2\pi)^3} \frac{ \gep_{mr}\gep_{ln}(p^2 \eta^{rl}-p^r p^l) \e^{-\ii p\cdot x} }{[p^+]^2(-p^2)^{2-\frac{d}{2}}}
\ee  

looks quite ugly. We indeed use the Schwinger formula

\be
\frac{1}{A^n} = \frac{\ii^{-n}}{\Gg(n)} \int_0^\infty \dd\ga\, \ga^{n-1} \e^{\ii \ga A}
\ee

and write (\ref{eq:uglyFT}) above as (up to the constant factor)

\be
\label{eq:moreuglyFT}
\begin{split}
& \int_{-\infty}^{+\infty} \frac{\dd^3 p}{(2\pi)^3} \frac{ (p^2 \eta^{rl}-p^r p^l) \e^{-\ii p\cdot x} }{[p^+]^2(-p^2)^{2-\frac{d}{2}}} \\
=&  (\d^r \d^l -\eta^{rl} \d^2)\int_{-\infty}^{+\infty} \frac{\dd^3 p}{(2\pi)^3}  \int_0^\infty \dd\ga\, \frac{1}{\sqrt{\ii\pi\ga}} \frac{\e^{\ii \ga (p^2+\ii\gep)} }{[p^+]^2}\\
=&  (\d^r \d^l -\eta^{rl} \d^2) \int_0^\infty \dd\ga\,  \frac{1}{2\ii\pi\ga} \e^{\frac{\ii}{4\ga}(x^T)^2} \, \int \frac{\dd p^+ \dd p^-}{(2\pi)^2} \, \frac{\e^{2\ii\ga p^+ p^- -\ii x^- p^+ -\ii x^+ p^- -\ga\gep} }{(p^+ + \ii\gep \,{\rm sign}(p^-))^2}
\end{split}
\ee  

where we have first inverted the Fourier and Schwinger integrals, implicitly assuming that the latter is convergent, as it will prove to be, and than we performed the straightforward integration over $p^T$. Of the two integrations which are left, the first is easily accomplished  with the help of a suitable prescription that guarantees convergence at $\pm\infty$

\be
\begin{split}
I_+ &=\lim_{\gd\to 0}\,\int_0^\infty \frac{\dd p^-}{2\pi} \frac{\e^{(2\ii\ga p^+  -\ii x^+ - \gd) p^- -\ii x^- p^+ -\ga\gep} }{(p^+ + \ii\gep)^2} = -\lim_{\gd\to 0}\, \frac{1}{2\pi} \frac{\e^{ -\ii x^- p^+ -\ga\gep} }{(2\ii\ga p^+  -\ii x^+ - \gd) (p^+ + \ii\gep)^2} \\
I_-&=\lim_{\gd\to 0}\,\int_{-\infty}^0 \frac{\dd p^-}{2\pi} \frac{\e^{(2\ii\ga p^+  -\ii x^+ + \gd) p^- -\ii x^- p^+ -\ga\gep} }{(p^+ - \ii\gep)^2} = \lim_{\gd\to 0}\, \frac{1}{2\pi} \frac{\e^{ -\ii x^- p^+ -\ga\gep} }{(2\ii\ga p^+  -\ii x^+ + \gd) (p^+ - \ii\gep)^2} 
\end{split}
\ee

where $\gd$ is real and positive. We are then left with only one integration to perform, the one over $p^+$. Note the pole structure of $I_+$ and $I_-$, as depicted in Figure \ref{fig:p+intcontours}. For $x^- >0$ the integral of $I_+$ converges on the lower half complex $p^+$ plane, where it has a double pole in $p^+ = -\ii\gep$ and a single pole in $p^+ = \frac{x^+-\ii\gd}{2\ga}$, Figure \ref{fig:p+intcontours} (a). On the other hand, in case $x^- <0$, the integral converges on the upper half plane, where it has no singularity, and is hence null. A totally analogous reasoning holds for the integral of $I_-$, up to the fact that its singularities are now placed on the upper half $p^+$ plane in a symmetric position, Figure \ref{fig:p+intcontours} (b). The first integral then reads

\be
\begin{split}
 \int_{-\infty}^{+\infty} \frac{\dd p^+}{2\pi} I_+ &=\frac{\ii \gth(x^-)}{2\pi} \left[{\rm Res}(I_+,-\ii\gep) + {\rm Res}(I_+, x^+-\ii\gd/2\ga) \right]\\
&= \frac{\ii \gth(x^-)}{2\pi} \left\{ \frac{x^-}{x^+ -\ii \gd -2\ii\ga\gep} + \frac{ 2\ga\ii\left( 1-\e^{-\ii\frac{x^+ x^-}{2\ga}} \right)  }{(x^+ -\ii \gd -2\ii\ga\gep)^2} \right\} \e^{-\ga\gep}
\end{split}
\ee

and in the same way one has

\be
 \int_{-\infty}^{+\infty} \frac{\dd p^+}{2\pi} I_- = \frac{\ii \gth(-x^-)}{2\pi} \left\{ \frac{x^-}{x^+ +\ii \gd +2\ii\ga\gep} + \frac{ 2\ga\ii\left( 1-\e^{-\ii\frac{x^+ x^-}{2\ga}} \right)  }{(x^+ +\ii \gd +2\ii\ga\gep)^2} \right\} \e^{-\ga\gep}
\ee

Note the amusing circumstance that the sign of the imaginary displacement of $x^+$ that guarantees convergence for both integrals above,  coherently reproduces the Mandelstam-Leibbrandt prescription of spurious poles in coordinate space $x^+\to x^+ -\ii\gd \,{\rm sign}(x^-)$. Putting all the bits together we are left with just one overall integral in the Schwinger parameter $\ga$

\be
\frac{1}{(2\pi)^2}\int_0^\infty \dd\ga\, \frac{\e^{\frac{\ii}{4\ga}(x^T)^2-\ga\gep}}{\ga} \left[ \frac{x^-}{[x^+]} + \frac{ 2\ga\ii\left( 1-\e^{-\ii\frac{x^+ x^-}{2\ga}} \right)  }{[x^+]^2} \right]
\ee

Again, the correct and self-consistent prescription of genuine and spurious poles in momentum space turns out to be essential. As a matter of fact, note that the integral above is regularized out-of-the-box. All integrations can be now performed exactly in terms of Bessel functions, but we are just interested in the small $\gep$ behaviour of the result

\be
A_1 = \int_0^\infty \dd\ga\, \frac{\e^{\frac{\ii}{4\ga}(x^T)^2-\ga\gep}}{\ga} \frac{x^-}{[x^+]} = - \frac{x^-}{[x^+]} \left[ 2\gg_E + \log\gep +\log\left(-\ii\frac{(x^T)^2}{4}\right)  \right]
\ee

\be
A_2= \int_0^\infty \dd\ga\, \frac{ 2\ii \e^{\frac{\ii}{4\ga}(x^T)^2} }{[x^+]^2}\, e^{-\ga\gep} = \frac{2\ii}{[x^+]^2} \left[ \frac{1}{\gep}  -\ii\frac{(x^T)^2}{4} \left( 2\gg_E -1 + \log\gep +\log\left(-\ii\frac{(x^T)^2}{4}\right) \right) \right]
\ee

\be
A_3= -\int_0^\infty \dd\ga\, \frac{ 2\ii\e^{-\ii\frac{x^2}{4\ga}} }{[x^+]^2}\, e^{-\ga\gep} = -\frac{2\ii}{[x^+]^2} \left[ \frac{1}{\gep}  +\ii\frac{x^2}{4} \left( 2\gg_E-1 + \log\gep +\log\left(\ii\frac{x^2}{4}\right) \right) \right]
\ee

As we claimed in equation (\ref{eq:moreuglyFT}) all divergences cancel and the result is finite

\be
\label{eq:alphaintegratedprop}
K(x)=A_1+A_2+A_3 = -\frac{x^-}{[x^+]} + \half \frac{x^2}{[x^+]} \log\left(-\frac{x^2}{(x^T)^2}\right)
\ee

At this point we are forced to compute explicitly the action of the derivatives in (\ref{eq:moreuglyFT}) on (\ref{eq:alphaintegratedprop}) (this is one of the drawback of axial gauges). Keeping in mind that all indices take values $+,T$ one finds

\be
\begin{split}
\label{eq:derivstructure}
D_{mn}= \gep_{mr}\gep_{ln} (\eta^{rl} \d^2 - \d^r \d^l ) & =(\gd_{ml}\gd_{rn} - \gd_{mn}\gd_{rl})(\eta^{rl} \d^2 - \d^r \d^l )\\
&= \d^2 \eta_{mn} - \d_m \d_n -\gd_{mn}(\d^2 \eta^{rr}-\d^r\d^r)\\
&= \d^2 \eta_{mn} - \d_m \d_n +\gd_{mn}(2\d^+\d^- +\d^+\d^+) 			
\end{split}
\ee

where we used the fact that $\Tr\eta=-1$ and $\d^2=2\d^+\d^- -\d^{T2}$. Hence we can finally write the gauge field one-loop corrected propagator in strictly three dimensions

\be
\label{eq:lcgauge1loopprop}
G^{(2)\,ab}_{mn} = \left( \frac{2\pi}{k} \right)^2 N \,\gd^I_I D_{mn}\left[ -\frac{x^-}{[x^+]} + \half \frac{x^2}{[x^+]^2} \log\left(-\frac{x^2}{(x^T)^2}\right) \right]
\ee

For future use, in the followings  we list the action of double derivatives on the kernel $K(x)$ of (\ref{eq:lcgauge1loopprop}).

\be
\label{eq:derivatives}
\d^+ K(x) = \frac{1}{[x^+]}\log\left( -\frac{x^2}{(x^T)^2} \right)
\ee

\be
\d^+\d^+ K(x) = \frac{\d}{\d x^-}\frac{\d}{\d x^-} K(x) = \frac{2}{x^2}
\ee

\be
\d^+\d^- K(x) = \frac{\d}{\d x^-}\frac{\d}{\d x^+} K(x) = \frac{1}{[x^+]^2} \log\left(-\frac{x^2}{(x^T)^2}\right) + \frac{2 x^- }{x^2 x^+}
\ee

\be
\d^+\d^T = -\frac{\d}{\d x^T}\frac{\d}{\d x^-} K(x) = \frac{4 x^-}{x^2 x^T}
\ee



\newpage
\bibliographystyle{JHEP}
\bibliography{biblio}

\providecommand{\href}[2]{#2}\begingroup\raggedright\begin{thebibliography}{10}

\bibitem{Korchemsky:1987wg}
G.~Korchemsky and A.~Radyushkin, {\it {Renormalization of the Wilson Loops
  Beyond the Leading Order}},  {\em Nucl.Phys.} {\bf B283} (1987) 342--364.

\bibitem{Korchemsky:1991zp}
G.~Korchemsky and A.~Radyushkin, {\it {Infrared factorization, Wilson lines and
  the heavy quark limit}},  {\em Phys.Lett.} {\bf B279} (1992) 359--366,
  [\href{http://xxx.lanl.gov/abs/hep-ph/9203222}{{\tt hep-ph/9203222}}].

\bibitem{Korchemsky:1988si}
G.~Korchemsky, {\it {Asymptotics of the Altarelli-Parisi-Lipatov Evolution
  Kernels of Parton Distributions}},  {\em Mod.Phys.Lett.} {\bf A4} (1989)
  1257--1276.

\bibitem{Balitsky:1987bk}
I.~Balitsky and V.~M. Braun, {\it {Evolution Equations for QCD String
  Operators}},  {\em Nucl.Phys.} {\bf B311} (1989) 541--584.

\bibitem{Korchemsky:1992xv}
G.~Korchemsky and G.~Marchesini, {\it {Structure function for large x and
  renormalization of Wilson loop}},  {\em Nucl.Phys.} {\bf B406} (1993)
  225--258, [\href{http://xxx.lanl.gov/abs/hep-ph/9210281}{{\tt
  hep-ph/9210281}}].

\bibitem{Craigie:1980qs}
N.~Craigie and H.~Dorn, {\it {ON THE RENORMALIZATION AND SHORT DISTANCE
  PROPERTIES OF HADRONIC OPERATORS IN QCD}},  {\em Nucl.Phys.} {\bf B185}
  (1981) 204.

\bibitem{Gubser:2002tv}
S.~Gubser, I.~Klebanov, and A.~M. Polyakov, {\it {A Semiclassical limit of the
  gauge / string correspondence}},  {\em Nucl.Phys.} {\bf B636} (2002) 99--114,
  [\href{http://xxx.lanl.gov/abs/hep-th/0204051}{{\tt hep-th/0204051}}].

\bibitem{Frolov:2002av}
S.~Frolov and A.~A. Tseytlin, {\it {Semiclassical quantization of rotating
  superstring in AdS(5) x S**5}},  {\em JHEP} {\bf 0206} (2002) 007,
  [\href{http://xxx.lanl.gov/abs/hep-th/0204226}{{\tt hep-th/0204226}}].

\bibitem{Beisert:2006ez}
N.~Beisert, B.~Eden, and M.~Staudacher, {\it {Transcendentality and Crossing}},
   {\em J.Stat.Mech.} {\bf 0701} (2007) P01021,
  [\href{http://xxx.lanl.gov/abs/hep-th/0610251}{{\tt hep-th/0610251}}].

\bibitem{Rey:1998ik}
S.-J. Rey and J.-T. Yee, {\it {Macroscopic strings as heavy quarks in large N
  gauge theory and anti-de Sitter supergravity}},  {\em Eur.Phys.J.} {\bf C22}
  (2001) 379--394, [\href{http://xxx.lanl.gov/abs/hep-th/9803001}{{\tt
  hep-th/9803001}}].

\bibitem{Maldacena:1998im}
J.~M. Maldacena, {\it {Wilson loops in large N field theories}},  {\em
  Phys.Rev.Lett.} {\bf 80} (1998) 4859--4862,
  [\href{http://xxx.lanl.gov/abs/hep-th/9803002}{{\tt hep-th/9803002}}].

\bibitem{Berenstein:1998ij}
D.~E. Berenstein, R.~Corrado, W.~Fischler, and J.~M. Maldacena, {\it {The
  Operator product expansion for Wilson loops and surfaces in the large N
  limit}},  {\em Phys.Rev.} {\bf D59} (1999) 105023,
  [\href{http://xxx.lanl.gov/abs/hep-th/9809188}{{\tt hep-th/9809188}}].

\bibitem{Drukker:1999zq}
N.~Drukker, D.~J. Gross, and H.~Ooguri, {\it {Wilson loops and minimal
  surfaces}},  {\em Phys.Rev.} {\bf D60} (1999) 125006,
  [\href{http://xxx.lanl.gov/abs/hep-th/9904191}{{\tt hep-th/9904191}}].

\bibitem{Drukker:2011za}
N.~Drukker and V.~Forini, {\it {Generalized quark-antiquark potential at weak
  and strong coupling}},  {\em JHEP} {\bf 1106} (2011) 131,
  [\href{http://xxx.lanl.gov/abs/1105.5144}{{\tt arXiv:1105.5144}}].

\bibitem{Correa:2012nk}
D.~Correa, J.~Henn, J.~Maldacena, and A.~Sever, {\it {The cusp anomalous
  dimension at three loops and beyond}},  {\em JHEP} {\bf 1205} (2012) 098,
  [\href{http://xxx.lanl.gov/abs/1203.1019}{{\tt arXiv:1203.1019}}].

\bibitem{Correa:2012at}
D.~Correa, J.~Henn, J.~Maldacena, and A.~Sever, {\it {An exact formula for the
  radiation of a moving quark in N=4 super Yang Mills}},  {\em JHEP} {\bf 1206}
  (2012) 048, [\href{http://xxx.lanl.gov/abs/1202.4455}{{\tt
  arXiv:1202.4455}}].

\bibitem{Drukker:2012de}
N.~Drukker, {\it {Integrable Wilson loops}},
  \href{http://xxx.lanl.gov/abs/1203.1617}{{\tt arXiv:1203.1617}}.

\bibitem{Correa:2012hh}
D.~Correa, J.~Maldacena, and A.~Sever, {\it {The quark anti-quark potential and
  the cusp anomalous dimension from a TBA equation}},  {\em JHEP} {\bf 1208}
  (2012) 134, [\href{http://xxx.lanl.gov/abs/1203.1913}{{\tt
  arXiv:1203.1913}}].

\bibitem{Zarembo:2002an}
K.~Zarembo, {\it {Supersymmetric Wilson loops}},  {\em Nucl.Phys.} {\bf B643}
  (2002) 157--171, [\href{http://xxx.lanl.gov/abs/hep-th/0205160}{{\tt
  hep-th/0205160}}].

\bibitem{Aharony:2008ug}
O.~Aharony, O.~Bergman, D.~L. Jafferis, and J.~Maldacena, {\it {N=6
  superconformal Chern-Simons-matter theories, M2-branes and their gravity
  duals}},  {\em JHEP} {\bf 0810} (2008) 091,
  [\href{http://xxx.lanl.gov/abs/0806.1218}{{\tt arXiv:0806.1218}}].

\bibitem{Aharony:2008gk}
O.~Aharony, O.~Bergman, and D.~L. Jafferis, {\it {Fractional M2-branes}},  {\em
  JHEP} {\bf 0811} (2008) 043, [\href{http://xxx.lanl.gov/abs/0807.4924}{{\tt
  arXiv:0807.4924}}].

\bibitem{Griguolo:2012iq}
L.~Griguolo, D.~Marmiroli, G.~Martelloni, and D.~Seminara, {\it {The
  generalized cusp in ABJ(M) N = 6 Super Chern-Simons theories}},
  \href{http://xxx.lanl.gov/abs/1208.5766}{{\tt arXiv:1208.5766}}.

\bibitem{Forini:2012bb}
V.~Forini, V.~G.~M. Puletti, and O.~Ohlsson~Sax, {\it {Generalized cusp in
  $AdS_4 x CP^3$ and more one-loop results from semiclassical strings}},
  \href{http://xxx.lanl.gov/abs/1204.3302}{{\tt arXiv:1204.3302}}.

\bibitem{Drukker:2009hy}
N.~Drukker and D.~Trancanelli, {\it {A Supermatrix model for N=6 super
  Chern-Simons-matter theory}},  {\em JHEP} {\bf 1002} (2010) 058,
  [\href{http://xxx.lanl.gov/abs/0912.3006}{{\tt arXiv:0912.3006}}].

\bibitem{Henn:2010ps}
J.~M. Henn, J.~Plefka, and K.~Wiegandt, {\it {Light-like polygonal Wilson loops
  in 3d Chern-Simons and ABJM theory}},  {\em JHEP} {\bf 1008} (2010) 032,
  [\href{http://xxx.lanl.gov/abs/1004.0226}{{\tt arXiv:1004.0226}}].

\bibitem{Chen:2011vv}
W.-M. Chen and Y.-t. Huang, {\it {Dualities for Loop Amplitudes of N=6
  Chern-Simons Matter Theory}},  {\em JHEP} {\bf 1111} (2011) 057,
  [\href{http://xxx.lanl.gov/abs/1107.2710}{{\tt arXiv:1107.2710}}].

\bibitem{Bianchi:2011dg}
M.~S. Bianchi, M.~Leoni, A.~Mauri, S.~Penati, and A.~Santambrogio, {\it
  {Scattering Amplitudes/Wilson Loop Duality In ABJM Theory}},
  \href{http://xxx.lanl.gov/abs/1107.3139}{{\tt arXiv:1107.3139}}.

\bibitem{Bianchi:2011fc}
M.~S. Bianchi, M.~Leoni, A.~Mauri, S.~Penati, and A.~Santambrogio, {\it
  {Scattering in ABJ theories}},  {\em JHEP} {\bf 1112} (2011) 073,
  [\href{http://xxx.lanl.gov/abs/1110.0738}{{\tt arXiv:1110.0738}}].

\bibitem{Kapustin:2009kz}
A.~Kapustin, B.~Willett, and I.~Yaakov, {\it {Exact Results for Wilson Loops in
  Superconformal Chern-Simons Theories with Matter}},  {\em JHEP} {\bf 1003}
  (2010) 089, [\href{http://xxx.lanl.gov/abs/0909.4559}{{\tt
  arXiv:0909.4559}}].

\bibitem{Marino:2009jd}
M.~Marino and P.~Putrov, {\it {Exact Results in ABJM Theory from Topological
  Strings}},  {\em JHEP} {\bf 1006} (2010) 011,
  [\href{http://xxx.lanl.gov/abs/0912.3074}{{\tt arXiv:0912.3074}}].

\bibitem{Drukker:2010nc}
N.~Drukker, M.~Marino, and P.~Putrov, {\it {From weak to strong coupling in
  ABJM theory}},  {\em Commun.Math.Phys.} {\bf 306} (2011) 511--563,
  [\href{http://xxx.lanl.gov/abs/1007.3837}{{\tt arXiv:1007.3837}}].

\bibitem{Drukker:2011zy}
N.~Drukker, M.~Marino, and P.~Putrov, {\it {Nonperturbative aspects of ABJM
  theory}},  \href{http://xxx.lanl.gov/abs/1103.4844}{{\tt arXiv:1103.4844}}.

\bibitem{Marino:2011eh}
M.~Marino and P.~Putrov, {\it {ABJM theory as a Fermi gas}},  {\em
  J.Stat.Mech.} {\bf 1203} (2012) P03001,
  [\href{http://xxx.lanl.gov/abs/1110.4066}{{\tt arXiv:1110.4066}}].

\bibitem{Cardinali:2012ru}
V.~Cardinali, L.~Griguolo, G.~Martelloni, and D.~Seminara, {\it {New
  supersymmetric Wilson loops in ABJ(M) theories}},
  \href{http://xxx.lanl.gov/abs/1209.4032}{{\tt arXiv:1209.4032}}.

\bibitem{Salpeter:1951sz}
E.~Salpeter and H.~Bethe, {\it {A Relativistic equation for bound state
  problems}},  {\em Phys.Rev.} {\bf 84} (1951) 1232--1242.

\bibitem{Erickson:1999qv}
J.~Erickson, G.~Semenoff, R.~Szabo, and K.~Zarembo, {\it {Static potential in
  N=4 supersymmetric Yang-Mills theory}},  {\em Phys.Rev.} {\bf D61} (2000)
  105006, [\href{http://xxx.lanl.gov/abs/hep-th/9911088}{{\tt
  hep-th/9911088}}].

\bibitem{Erickson:2000af}
J.~Erickson, G.~Semenoff, and K.~Zarembo, {\it {Wilson loops in N=4
  supersymmetric Yang-Mills theory}},  {\em Nucl.Phys.} {\bf B582} (2000)
  155--175, [\href{http://xxx.lanl.gov/abs/hep-th/0003055}{{\tt
  hep-th/0003055}}].

\bibitem{Makeenko:2006ds}
Y.~Makeenko, P.~Olesen, and G.~W. Semenoff, {\it {Cusped SYM Wilson loop at two
  loops and beyond}},  {\em Nucl.Phys.} {\bf B748} (2006) 170--199,
  [\href{http://xxx.lanl.gov/abs/hep-th/0602100}{{\tt hep-th/0602100}}].

\bibitem{Bykov:2012sc}
D.~Bykov and K.~Zarembo, {\it {Ladders for Wilson Loops Beyond Leading Order}},
   {\em JHEP} {\bf 1209} (2012) 057,
  [\href{http://xxx.lanl.gov/abs/1206.7117}{{\tt arXiv:1206.7117}}].

\bibitem{Frohlich:1989gr}
J.~Frohlich and C.~King, {\it {THE CHERN-SIMONS THEORY AND KNOT POLYNOMIALS}},
  {\em Commun.Math.Phys.} {\bf 126} (1989) 167.

\bibitem{AlvarezGaume:1989wk}
L.~Alvarez-Gaume, J.~Labastida, and A.~Ramallo, {\it {A Note on Perturbative
  Chern-Simons Theory}},  {\em Nucl.Phys.} {\bf B334} (1990) 103.

\bibitem{Alvarez:1991sx}
M.~Alvarez and J.~Labastida, {\it {Analysis of observables in Chern-Simons
  perturbation theory}},  {\em Nucl.Phys.} {\bf B395} (1993) 198--238,
  [\href{http://xxx.lanl.gov/abs/hep-th/9110069}{{\tt hep-th/9110069}}].

\bibitem{Labastida:1997uw}
J.~Labastida and E.~Perez, {\it {Kontsevich integral for Vassiliev invariants
  from Chern-Simons perturbation theory in the light cone gauge}},  {\em
  J.Math.Phys.} {\bf 39} (1998) 5183--5198,
  [\href{http://xxx.lanl.gov/abs/hep-th/9710176}{{\tt hep-th/9710176}}].

\bibitem{Bassetto:1993xd}
A.~Bassetto, I.~Korchemskaya, G.~Korchemsky, and G.~Nardelli, {\it {Gauge
  invariance and anomalous dimensions of a light cone Wilson loop in lightlike
  axial gauge}},  {\em Nucl.Phys.} {\bf B408} (1993) 62--90,
  [\href{http://xxx.lanl.gov/abs/hep-ph/9303314}{{\tt hep-ph/9303314}}].

\bibitem{Brodsky:1997de}
S.~J. Brodsky, H.-C. Pauli, and S.~S. Pinsky, {\it {Quantum chromodynamics and
  other field theories on the light cone}},  {\em Phys.Rept.} {\bf 301} (1998)
  299--486, [\href{http://xxx.lanl.gov/abs/hep-ph/9705477}{{\tt
  hep-ph/9705477}}].

\bibitem{Drukker:2006zk}
N.~Drukker, S.~Giombi, R.~Ricci, and D.~Trancanelli, {\it {On the D3-brane
  description of some 1/4 BPS Wilson loops}},  {\em JHEP} {\bf 0704} (2007)
  008, [\href{http://xxx.lanl.gov/abs/hep-th/0612168}{{\tt hep-th/0612168}}].

\bibitem{Drukker:2008zx}
N.~Drukker, J.~Plefka, and D.~Young, {\it {Wilson loops in 3-dimensional N=6
  supersymmetric Chern-Simons Theory and their string theory duals}},  {\em
  JHEP} {\bf 0811} (2008) 019, [\href{http://xxx.lanl.gov/abs/0809.2787}{{\tt
  arXiv:0809.2787}}].

\bibitem{Rey:2008bh}
S.-J. Rey, T.~Suyama, and S.~Yamaguchi, {\it {Wilson Loops in Superconformal
  Chern-Simons Theory and Fundamental Strings in Anti-de Sitter Supergravity
  Dual}},  {\em JHEP} {\bf 0903} (2009) 127,
  [\href{http://xxx.lanl.gov/abs/0809.3786}{{\tt arXiv:0809.3786}}].

\bibitem{Kruczenski:2002fb}
M.~Kruczenski, {\it {A Note on twist two operators in N=4 SYM and Wilson loops
  in Minkowski signature}},  {\em JHEP} {\bf 0212} (2002) 024,
  [\href{http://xxx.lanl.gov/abs/hep-th/0210115}{{\tt hep-th/0210115}}].

\bibitem{Beisert:2005tm}
N.~Beisert, {\it {The SU(2|2) dynamic S-matrix}},  {\em Adv.Theor.Math.Phys.}
  {\bf 12} (2008) 945--979, [\href{http://xxx.lanl.gov/abs/hep-th/0511082}{{\tt
  hep-th/0511082}}].

\bibitem{Grignani:2008is}
G.~Grignani, T.~Harmark, and M.~Orselli, {\it {The SU(2) x SU(2) sector in the
  string dual of N=6 superconformal Chern-Simons theory}},  {\em Nucl.Phys.}
  {\bf B810} (2009) 115--134, [\href{http://xxx.lanl.gov/abs/0806.4959}{{\tt
  arXiv:0806.4959}}].

\bibitem{Nishioka:2008gz}
T.~Nishioka and T.~Takayanagi, {\it {On Type IIA Penrose Limit and N=6
  Chern-Simons Theories}},  {\em JHEP} {\bf 0808} (2008) 001,
  [\href{http://xxx.lanl.gov/abs/0806.3391}{{\tt arXiv:0806.3391}}].

\bibitem{Berenstein:2008dc}
D.~Berenstein and D.~Trancanelli, {\it {Three-dimensional N=6 SCFT's and their
  membrane dynamics}},  {\em Phys.Rev.} {\bf D78} (2008) 106009,
  [\href{http://xxx.lanl.gov/abs/0808.2503}{{\tt arXiv:0808.2503}}].

\bibitem{Berenstein:2009qd}
D.~Berenstein and D.~Trancanelli, {\it {S-duality and the giant magnon
  dispersion relation}},  \href{http://xxx.lanl.gov/abs/0904.0444}{{\tt
  arXiv:0904.0444}}.

\bibitem{Minahan:2008hf}
J.~Minahan and K.~Zarembo, {\it {The Bethe ansatz for superconformal
  Chern-Simons}},  {\em JHEP} {\bf 0809} (2008) 040,
  [\href{http://xxx.lanl.gov/abs/0806.3951}{{\tt arXiv:0806.3951}}].

\bibitem{Gaiotto:2008cg}
D.~Gaiotto, S.~Giombi, and X.~Yin, {\it {Spin Chains in N=6 Superconformal
  Chern-Simons-Matter Theory}},  {\em JHEP} {\bf 0904} (2009) 066,
  [\href{http://xxx.lanl.gov/abs/0806.4589}{{\tt arXiv:0806.4589}}].

\bibitem{Guadagnini:1989kr}
E.~Guadagnini, M.~Martellini, and M.~Mintchev, {\it {Perturbative Aspects of
  the Chern-Simons Field Theory}},  {\em Phys.Lett.} {\bf B227} (1989) 111.

\bibitem{Giavarini:1992xz}
G.~Giavarini, C.~Martin, and F.~Ruiz~Ruiz, {\it {Chern-Simons theory as the
  large mass limit of topologically massive Yang-Mills theory}},  {\em
  Nucl.Phys.} {\bf B381} (1992) 222--280,
  [\href{http://xxx.lanl.gov/abs/hep-th/9206007}{{\tt hep-th/9206007}}].

\bibitem{Witten:1988hf}
E.~Witten, {\it {Quantum Field Theory and the Jones Polynomial}},  {\em
  Commun.Math.Phys.} {\bf 121} (1989) 351.

\bibitem{Leibbrandt:1991qr}
G.~Leibbrandt and C.~Martin, {\it {Perturbative Chern-Simons theory in the
  light cone gauge: The One loop vacuum polarization tensor in a gauge
  invariant formalism}},  {\em Nucl.Phys.} {\bf B377} (1992) 593--621.

\bibitem{Leibbrandt:1993mr}
G.~Leibbrandt and C.~Martin, {\it {The Light cone gauge, Chern-Simons theory
  and topologically massive Yang-Mills theory}},  {\em Nucl.Phys.} {\bf B416}
  (1994) 351--376.

\bibitem{Mandelstam:1982cb}
S.~Mandelstam, {\it {Light Cone Superspace and the Ultraviolet Finiteness of
  the N=4 Model}},  {\em Nucl.Phys.} {\bf B213} (1983) 149--168.

\bibitem{Leibbrandt:1983pj}
G.~Leibbrandt, {\it {The Light Cone Gauge in Yang-Mills Theory}},  {\em
  Phys.Rev.} {\bf D29} (1984) 1699.

\bibitem{Bassetto:1992gv}
A.~Bassetto, {\it {The Free vector propagator in the light cone gauge and the
  Mandelstam-Leibbrandt prescription}},  {\em Phys.Rev.} {\bf D46} (1992)
  3676--3677.

\bibitem{Bassetto:1984dq}
A.~Bassetto, M.~Dalbosco, I.~Lazzizzera, and R.~Soldati, {\it {Yang-Mills
  Theories in the Light Cone Gauge}},  {\em Phys.Rev.} {\bf D31} (1985) 2012.

\end{thebibliography}\endgroup

\end{document}